\documentclass[twocolumn,amsmath,amssymb,superscriptaddress]{revtex4}
\usepackage{amsmath,amssymb,amsfonts}
\usepackage{graphics,subfigure}
\usepackage{epsfig}
\usepackage{color}
\usepackage{graphicx}

\newcommand{\kmin}{k_{\textrm{min}}}

\renewcommand{\eqref}[1]{Eq.~(\ref{#1})}

\newcommand{\pairmodel}{pair model }
\newcommand{\bachelormodel}{bachelor model }
\newcommand{\pairmodelpunt}{pair model}
\newcommand{\bachelormodelpunt}{bachelor model}

\newcommand{\avk}{\langle k \rangle}
\newcommand{\average}[1]{\langle k^{#1} \rangle}

\newcommand{\twee}{\epsilon} 
\newcommand{\avknn}{\langle k_{n}\rangle}

\newcommand{\ef}[1]{f_{#1}}
\newcommand{\isarray}{&=}
\newcommand{\barabasi}{Barab\'{a}si }
\newcommand{\barabasialbert}{Barab\'{a}si-Albert }
\newcommand{\erdosrenyi}{Erd\H{o}s-R\'{e}nyi }

\newcommand{\nearne}{nearest-neigbour }

\begin{document}

\title{Degree-dependent network growth: From preferential attachment to explosive percolation}
\author{Hans Hooyberghs}
\email{hans@itf.fys.kuleuven.be}
\affiliation{Instituut voor Theoretische Fysica, KU Leuven, Celestijnenlaan 200D,  B-3001 Leuven, Belgium}
\affiliation{Flemish Institute for Technological Research (VITO), Boerentang 200, B-2400 Mol, Belgium}
\author{Bert Van Schaeybroeck}
\affiliation{ Koninklijk Metereologisch Instituut (KMI), Ringlaan 3,
B-1180 Brussels, Belgium}
\author{Joseph O. Indekeu}
\affiliation{Instituut voor Theoretische Fysica, KU Leuven, Celestijnenlaan 200D,  B-3001 Leuven, Belgium}

\begin{abstract}
We present a simple model of network growth and solve it by writing down the dynamic equations for its macroscopic characteristics like the degree distribution and degree correlations. This allows us to study carefully the percolation transition using a generating functions theory. The model considers a network with a fixed number of nodes  wherein links are introduced using degree-dependent linking probabilities $p_k$.
To illustrate the techniques and support our findings using Monte-Carlo simulations, we introduce the exemplary linking rule $p_k\propto k^{-\alpha}$, with $\alpha$ between $-1$ and $+\infty$. This parameter may be used to interpolate between different regimes. For negative $\alpha$, links are most likely  attached to high-degree nodes. On the other hand, in case $\alpha>0$, nodes with low degrees are connected and the model asymptotically approaches a process undergoing explosive percolation. 
\end{abstract}
\maketitle

\section{Introduction}
An important research topic in network theory concerns the way in which networks grow. A few decades ago,  network construction models almost exclusively used random link addition. In the meantime,  a paradigm shift occurred, influenced by the realisation that the growth of real-life complex systems can be better described by a process  which exploits specific linking rules. The inaugural paper of Barab\'{a}si and Albert was the first to capture two key ingredients of network growth in real-life structures: continuous growth and preferential attachment \cite{barabasi1999}. 
Not all construction models, however, aim at introducing the most realistic description of real-life processes. Instead, some network growth models are designed  to study the effect of specific linking rules on the evolution of the network structure. In this paper, we are particularly interested in the influence on the properties of  the formation of a giant connected component in the network, the so-called percolation transition \cite{stauffer1994}. In recent years, much attention was devoted to this transition in complex networks, which was, amongst other, studied as a model for virus spreading \cite{kenah2007} and network failures \cite{cohen2000,dorogovtsev2007,moreira2009,hooyberghs2010biased,hooyberghs2010bipartite}.

Recently, Achlioptas \textit{et al.}~discovered a new type of percolation transition in complex networks \cite{achlioptas2009}. The transition which they coined \textit{explosive} is marked by an abrupt growth of the largest cluster in large networks. Following the ground-breaking work of Achlioptas \textit{et al.}, explosive transitions have been observed and studied in a wide variety of network and lattice models  \cite{ziff2009,radicchi2009,moreira2010,friedman2009,hooyberghs2011explosive}. The exact nature of the transition has been the subject of a vivid debate. By now, it has been established that the transition caused by the original process of Achlioptas \textit{et al.}~is continuous~\cite{costa2010,grassberger2011,riordan2011,lee2011}, while first-order transitions could be observed in models with global competition between the links \cite{nagler2011}. For instance, the discontinuity of a process in which clusters are randomly connected has been analytically and numerically shown in Refs.~\cite{cho2010,manna2011}.

We introduce a model which studies the growth of a network with a fixed number of nodes using general degree-dependent linking probabilities. The main properties of the constructed networks and the percolation transition are studied. One of the major advantages of this model is its analytical ease of handling. Numerical results to exact rate equations provide expressions for the degree distribution and degree correlations in the constructed networks by means of which the percolation properties can be studied accurately within a generating functions formalism.
To illustrate the techniques and support the findings using Monte Carlo simulations, we introduce two examples of linking rules. In both variants, a single parameter $\alpha$ determines the dynamics. The parameter may be used to interpolate between different regimes: depending on the sign of $\alpha$ new edges are preferably attached to nodes with larger or smaller degrees. The limiting values of $\alpha$ correspond on the one hand to the attachment rule of Barab\'{a}si and Albert, and, on the other hand, to an explosive process that is similar to the ones reported in Refs.~\cite{cho2010,manna2011}.

The outline of this Paper is as follows.  In the next section, we introduce our network growth process and compare it with existing models.  In the remainder, the most prominent properties of the growth process and the constructed networks are studied. In the third and fourth section, we study rate equations which determine the evolution of the degree distribution and degree correlations for general linking rules. For two exemplary linking rules, we compare the numerical solutions to this approach with simulations. Section \ref{sectionpercolation} provides a way to obtain the percolation properties using a generating functions scheme. Numerical results within the scheme are again compared with simulations for two exemplary linking rules. Throughout the Paper, we pay special attention to the parameter regimes for which explosive percolation, preferential attachment or random network growth are recovered. Expansions and particular analytical techniques provide deeper insight into these limiting situations.

\section{The model}
\subsection{Linking probabilities}
Consider a network consisting of $N$ sparsely connected nodes. As time evolves, links (edges) are added between the nodes. The probability to attach an edge to node $i$ depends only on its \textit{instantaneous} degree $k_i$. In general, we define the \emph{linking weights} $p_{k_i}$ as
\begin{equation}\label{prob}
p_{k_i} \equiv p_{k} = \frac{\ef{k_i}}{c(t)},
\end{equation}
with
\begin{equation}\label{ct}
c(t)=\frac{1}{N}\sum_{j=1}^N \ef{k_j}
\end{equation}
the time-dependent normalisation. 
Here $\ef{k}$ is an algebraic function of the degree $k$, which will be specified later on. Note that we define the linking weights such that $N=\sum_i p_{k_i}$ and thus $1=\sum_k p_k P(k)$, with $P(k)$ the degree distribution. In each time-step, two nodes are selected according to 
the probability of \eqref{prob} and a new link is introduced between them. Multiple links between the same nodes and self-linking are avoided. The linking procedure can continue until a fully-connected network is reached, but  we are mainly interested in the percolation transition, which happens much earlier in the linking process. Moreover, in the remainder of the text, we focus on the macroscopic limit, thus implicitly assuming $N\rightarrow\infty$. 

In most network growth models, both the number of links and nodes evolve dynamically. Our model is different in the sense that the number of nodes in the network is fixed.  Noteworthy models with constant number of nodes were presented by Barab\'{a}si et al.~\cite{barabasi1999b} and also by G\'omez-Garde\~nes and Moreno~\cite{gomez2006}. In their models, in each time-step a random  node is linked to a node chosen using degree-dependent linking rules. Our model is different from theirs since both nodes are chosen according to a probability function $p_k$. In Sect.~\ref{appendixassymetric}, we relax the requirement of symmetric linking probabilities and instead consider processes in which both endnodes are chosen with different linking probabilities. Note that the main results of this Paper can be extended to such asymmetric network growth models.

\subsection{The bachelor and the \pairmodel}
To illustrate our findings using simulations we introduce two examples of degree-dependent linking rules. In the first diversification, we focus on a model with the most simple initial condition, i.e., all nodes are isolated at the start of the linking procedure. The linking probability is defined by
\begin{equation}
\ef{k}^b = (k+1)^{-\alpha},
\end{equation}
with a single parameter $\alpha$ and with $k\geq \kmin^b$, where $\kmin^b\equiv 0$ denotes the minimal degree in the network. The specific form of the function $f^b_k$ ensures that the weights are always well defined. Since initially all nodes are isolated, we call this model the \textit{bachelor model}. 
In the second variant we assume that
\begin{equation}
\ef{k}^p = k^{-\alpha}.
\end{equation}
Since the above function is ill-defined for connectivity $k=0$, all nodes are initially combined into pairs and the minimal connectivity is therefore $\kmin^p\equiv 1$. The model is henceforth denoted as the \textit{\pairmodelpunt}. Links can be classified as either \textit{initial-time} or  \textit{finite-time} links. 

In both models, a single parameter $\alpha$ determines the dynamics. The value $\alpha = 0$ indicates random network growth amounting to an Erd\H{o}s-R\'{e}nyi network with a Poisson-like degree distribution \cite{erdos1960}. For positive $\alpha$, nodes with a small degree are most likely selected which prevents the formation of hubs. For negative $\alpha$, on the other hand, links are most likely attached to nodes with a large connectivity. A similar degree-dependent attachment is found in the Barab\'{a}si-Albert model and the extension thereof by Krapivsky \textit{et al.} \cite{barabasi1999,krapivsky2000,krapivsky2001}. In their models, nodes are continuously added to the network and connected in a degree-dependent manner with older nodes.  Krapivsky \textit{et al.}~used the same linking probability as in our \pairmodelpunt, $p_k \propto k^{-\alpha}$, while in the \barabasialbert model $\alpha = -1$ is applied. In the following sections, we discuss the differences between models with a fixed number of nodes and models in which the number of nodes evolves by comparing our model with those of  Krapivksy \textit{et al.}~and Barab\'{a}si \textit{et al.}

In the remainder, we limit $\alpha$ to the range $[-1, +\infty[$. If $\alpha <-1$, simulations indicate that a single node can capture a non-vanishing fraction of all the links. This special case is studied more in detail in Ref.~\cite{phdthesis}.
In the lower limit, $\alpha = -1$, we retrieve the link attachment rule of the Barab\'{a}si-Albert model \cite{barabasi1999}. In the other extreme case, $\alpha = + \infty$, a new link will always be laid between (two of) the least connected nodes in the network. If all nodes contain at least one and at most two links, this routine is identical to one in which links are laid between two end-nodes of two randomly chosen clusters. In Refs.~\cite{cho2010,manna2011} it was proven that this cluster-linking process gives rise to an explosive first-order percolation transition. We will study the transition to the explosive regime in more detail in Section \ref{sectionpercolation}.

In the remainder of the paper, the differences and similarities between both diversifications are discussed thoroughly. We use the convention that in case no superscript (b or p) is specified, results apply to both variants. Throughout this paper, numerical solutions to the analytical results are compared with Monte Carlo simulations for both exemplary linking rules. We always use networks with $10^6$ nodes and average over 100 realisations of the network growth.

\section{Degree distribution\label{sectionconstructiondegreedistribution}}
\subsection{Rate equation}
In our model, the time evolution of the degree distribution is described by a mean-field rate equation:
\begin{equation}\label{sam}
\frac{\partial}{\partial t}P(k) = p_{k-1}P(k-1) -  p_kP(k).
\end{equation}
The first term on the right-hand-side describes the gain in nodes with degree $k$ in case a node with degree $k-1$ is selected, while the second term quantifies the loss if a node with degree $k$ is chosen. The time $t$ is defined such that $t$ increases with $\Delta t=2/N$ each time a single link is laid. \eqref{sam} is exact in the macroscopic limit ($N \rightarrow \infty$) when $\Delta t\rightarrow 0$ \footnote{Note that, upon discretisation of the time derivative, \eqref{sam} will also be exact for finite networks when averaging over different identical realisations and upon exclusion of self-linking and multiple links between the same nodes. In practice, the possibility of self-linking is small at all stages of the linking procedure. If $\alpha \geq -1$, the probability to select a node twice at the same time is proportional to $N^{-1}$, which is negligible in the limit of an infinite network. The probability to add a second link between the same nodes is negligible during the early stages of the network growth. Simulations indicate that multiple linking does not occur if $\alpha \geq -1$ as long as $\avk\lesssim 35$. Since we are mostly interested in the percolation transition, which happens when $\avk \leq 2$, we neglect the possibility of multiple linking. }. A detailed derivation of the rate equation can be found in Appendix \ref{appendixrateeqdd}.

The rate equation can be directly integrated with numerical techniques. However, we continue now by solving it exactly. If $f_k \neq f_q$ for all $k\neq q$, its solution can be cast into the form:
\begin{equation}\label{ansatz1} 
 P(k,t) = \sum_{q=q_{min}}^kA(q,k)[x(t)]^{f_q},
\end{equation}
where we made explicit the time dependence of the degree distribution. The time dependence stems solely from the function $x(t)$, defined as
\begin{equation}\label{defx}
x(t) = \exp\left(-\int_0^t\frac{1}{c(t')}dt' \right).
\end{equation}
Substitution in the rate equation yields the time-independent constants,
\begin{equation}\label{A_coeff}
A(q,k) = \frac{\ef{q}}{\ef{k}}\prod_{m\neq q}^k\left(1-\frac{\ef{q}}{\ef{m}}\right)^{-1},
\end{equation}
while, due to the specific structure of \eqref{ansatz1}, $x(t)$ can be obtained without performing a numerical integration, by solving the differential equation
\begin{subequations}
\begin{align}
\frac{dx}{dt} &= -\frac{x(t)}{c(t)}\label{dvglx}
\end{align}
\end{subequations}
exactly. Inserting $c(t) = \sum_k \ef{k} P(k)$ and applying separation of variables yields
\begin{equation}\label{determinex1}
t = \sum_{k=q_{min}} \sum_{q=q_{min}}^k \frac{\ef{k}}{\ef{q}}A(q,k)  \left(1-[x(t)]^{f_q}\right).
\end{equation}
This implicit equation for the time-dependent part of the degree distribution can be solved numerically. This routine can be performed in each time-step separately with an arbitrary accuracy, thereby avoiding cumulative errors, unavoidably associated with the numerical scheme necessary to integrate \eqref{sam} directly.

\subsection{Application to the pair and the \bachelormodel}
The application of the general scheme developed in the previous section is illustrated using the bachelor model and the \pairmodelpunt. Recall that the initial conditions are different in both models: in the \bachelormodelpunt, all nodes are initially isolated, while all nodes are paired up for the \pairmodelpunt. For the degree distribution, this entails:
\begin{subequations}\label{init_models}
\begin{align}
P^b(k,t=0)=&\delta_{0,k}\quad \Rightarrow \quad\avk^b(t=0)=0 ,\label{single_model_init}\\
P^p(k,t=0)=&\delta_{1,k}\quad \Rightarrow \quad\avk^p(t=0)=1 .\label{pair_model_init}
\end{align}
\end{subequations}
Moreover, the time is directly related to the average degree since  $t = \avk^b$ for the \bachelormodelpunt, while $t=\avk^p-1$ for the \pairmodelpunt. Note finally that the procedure outlined in Eqs.~(\ref{ansatz1}) - (\ref{determinex1}) becomes ill-defined if $\alpha =0$, since the weights are constant in this special case. We will discuss this random network growth separately further on. 

For $\alpha \neq 0$, it is by inspection easily verified that
\begin{subequations}
\begin{align}
\ef{k}^b&= \ef{k+1}^p,\\
A^b(q,k)&= A^p(q+1,k+1),\label{A_cst_pair_single}\\
x^b(t)&=x^p(t),\\
c^b(t)&=c^p(t),
\end{align}
\end{subequations}
where the last two expressions are proven using \eqref{determinex1}. These relations are anticipated since 
 the evolution of the weights is the same in both models. Consequently, the time-dependent parts of the degree distributions are also equal, such that, at any time, a simple relation between the degree distributions of both models exists:
\begin{equation}\label{n_pair_single}
 P^p(k+1,t)=P^b(k,t).
\end{equation}
Thus, at any time, the probability to observe a node with degree $k+1$ in the \pairmodel is the same as the probability to observe a node with degree $k$ in the \bachelormodelpunt. 

\begin{figure}[t]
\centering
\includegraphics[angle=0,width = .5 \textwidth]{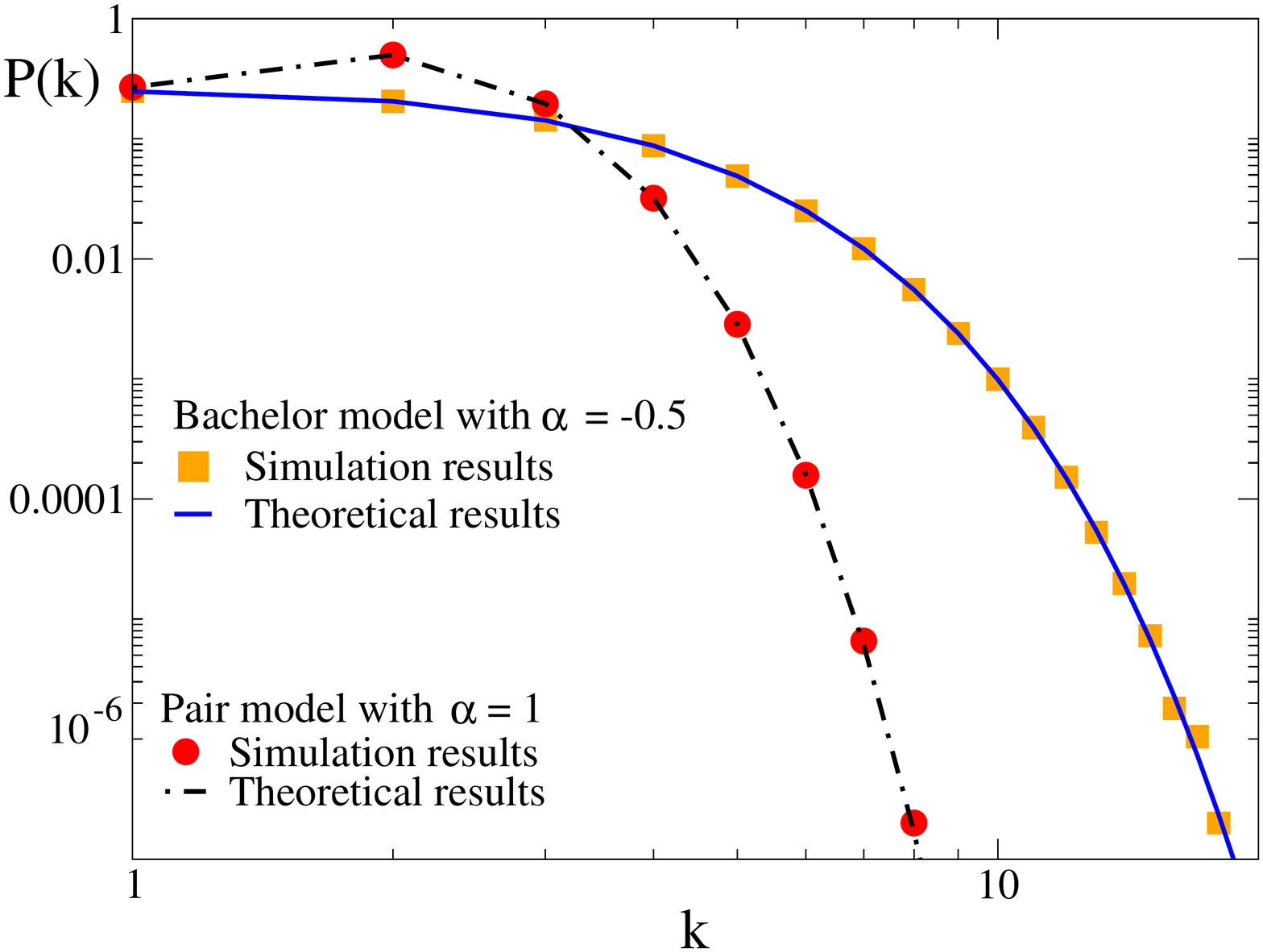}
\caption[ Degree distribution]{(Colour online) Comparison of analytical results (lines) and simulations (symbols) for the degree distributions $P^b(k,t=2)$ and $P^p(k,t=1)$ as a function of the degree $k$. For all cases $\avk = 2$. The Figure shows results for the \bachelormodel with $\alpha = -0.5$ (orange squares and blue line) and for the \pairmodel with $\alpha = 1$ (red dots and black dot-dashed line) on a logarithmic scale. All simulation data are averages over 100 realisations of the  growth of a network with $10^6$ nodes.  \label{degreedistr}}
\end{figure}

For both models, an excellent agreement between the numerical and the simulational degree distribution is found for a wide range of $\alpha$-values. As an example, Fig.~\ref{degreedistr} shows the results for the \pairmodel with $\alpha = 1$ and for the \bachelormodel with $\alpha = -0.5$, both for $\avk = 2$. 
Extensive studies of simulation data moreover indicate that, within the range  $\alpha\in{]-1,+\infty]\backslash\{0\}}$, a power-law degree distribution is never found and the constructed networks are thus not of the scale-free type.  Krapivsky \textit{et al.}~found similar results in their model. They could only construct a scale-free network for $\alpha=-1$, while for $-1 < \alpha < 0$ their networks comprised a stretched-exponential degree distribution \cite{krapivsky2001}. The special cases $\alpha=0$ and $\alpha = -1$ are solved analytically in the next subsection. 

\subsection{Special cases}
Although for general values of $\alpha$ the rate equations must be solved with numerical techniques, we can make progress analytically in some special cases. We focus on random link addition ($\alpha = 0$), preferential attachment ($\alpha = -1$) and the transition towards an explosive process ($\alpha\rightarrow\infty$). Note that these limiting situations capture three qualitatively distinct regimes. We also consider an asymmetric model in which $\alpha = 0$ is mixed with $\alpha = -1$. 

\subsubsection{Random link addition ($\alpha=0$)\label{alpha0}}
Upon solving \eqref{sam} in the previous subsection it was  assumed that $\ef{k}$ is not constant. In the special case $\alpha = 0$, these derivations therefore no longer hold and the constants $A(q,k)$ are ill-defined. Using a generating functions approach we can, however, proceed analytically in this special case.

 Introducing the generating function $\mathcal{F}(z,t) = \sum_k P(k,t) z^k$, the rate equation becomes
\begin{equation}
\frac{\partial \mathcal{F}}{\partial t}(z,t) = (z-1)\mathcal{F}(z,t).
\end{equation}
Solving for $\mathcal{F}(z,t)$ using the appropriate boundary conditions and subsequently expanding in factors of $z$ yields
\begin{align}
P^p(k+1,t)=P^b(k,t)=\frac{t^{k}}{k!}e^{-t}.\label{dda0}
\end{align}
We thus find that the relation between the degree distributions of both models, $ P^p(k+1,t)=P^b(k,t)$, which was derived for the case $\alpha\neq 0$, still holds if nodes are selected randomly. In \eqref{dda0} one recognises the Poisson distribution with mean $\avk$, which is a well-known result for Erd\H{o}s-R\'{e}nyi networks \cite{erdos1960,bollobas2001}.

\subsubsection{Preferential attachment ($\alpha = -1$)\label{alpha-1}}
Also in the limiting case $\alpha = -1$,  the degree distribution can be obtained analytically. Since the products in the definition of $A(q,k)$, 
\eqref{A_coeff}, can be computed analytically, the implicit equation  for $x(t)$, \eqref{determinex1}, takes  the simple form
\begin{equation}
t = \frac{1-x}{x}.
\end{equation}
The degree distribution is then readily derived as
\begin{align}
P^p(k+1,t)=P^b(k,t) = \frac{1}{1+t}\left(\frac{t}{1+t}\right)^{k}\label{dda-1}.
\end{align}
Simulations again confirm these analytic expressions. For both models, in each time-step, the degree distribution is of the form $P(k) \propto C^k$ with a (time-dependent) constant $C$. Although preferential attachment is present, the degree distribution is not of scale-free nature. A similar observation has been made by Barab\'{a}si \textit{et al.}, who argued that both  preferential attachment and  network growth are essential for the formation of a scale-free network \cite{barabasi1999b}. However, they used a model in which one endpoint of a new link is chosen using preferential attachment, while the other is chosen randomly. Our findings so far apply to a growth model with preferential attachment at both ends of a new link.

\newcommand{\head}{\textrm{head}}
\newcommand{\tail}{\textrm{tail}}
\subsubsection{Asymmetric linking process \label{appendixassymetric}}
In the previous two cases, both endnodes of a new link are selected using the same linking probabilities. However, with similar techniques, we can study asymmetric linking processes in which different weights are used for the head and the tail of the new link. For example, one could be interested in a process which starts from a network with $N$ initially isolated nodes in which the head of a new link is chosen using preferential attachment, while the tail is chosen randomly. Note that a similar process was previously studied by \barabasi \textit{et al.}~in Ref.~\cite{barabasi1999b}.

In a general asymmetric process, the head of a link is placed using linking probabilities
\begin{equation}
p_k^\head \equiv \frac{f_k^\head}{c^\head(t)},
\end{equation}
while the tail is laid using linking probabilities
\begin{equation}
p_k^\tail \equiv \frac{f_k^\tail}{c^\tail(t)},
\end{equation}
where $f_k^\head$ and $f_k^\tail$ are algebraic functions of the degree $k$, and $c^\head(t)$ and $c^\tail(t)$ are the time-dependent normalisations. For instance, for the process mentioned above, $f_k^\head = k+1$ and $f_k^\tail=1$. Note moreover that the distinction between the head and the tail of the link is arbitrary and that both can be interchanged. Instead of choosing one end using the probabilities for the head and the other using those for the tail, we could therefore equally well apply a stochastic hybrid process in which the weights of the head are used with probability $\omega$, while the weights of the tail are applied with probability $1-\omega$. As an extension of the example mentioned above, we could for instance opt to choose an endnode randomly with probability $\omega$, while preferential attachment is applied with probability $1-\omega$. In the following, the rate equations will be extended to general hybrid processes. For our specific combination of random growth and preferential attachment, the rate equation will be solved exactly.

Extensions of the rate equations for (non-deterministic) symmetric processes are straightforwardly derived. For instance, the rate equation for the degree distribution, \eqref{sam}, becomes
\begin{widetext}\label{samasymetric}
\begin{align}
\frac{\partial}{\partial t}P(k) &= \left(\omega p_{k-1}^\head+(1-\omega)p_{k-1}^\tail\right)P(k-1)-\left(\omega p_{k}^\head+(1-\omega)p_{k}^\tail\right)P(k).
\end{align}
\end{widetext}
Similarly, rate equations for the (next-)nearest-neighbour correlations can be deduced. For general linking rules, the rate equations can again be solved numerically, while, for some specific choices, we can proceed analytically. For instance, for the example mentioned above, the rate equation for the degree distribution becomes
\begin{widetext}
\begin{align}
\frac{\partial}{\partial t}P(k) &= \left(\omega+(1-\omega)\frac{k}{t+1}\right)P(k-1)- \left(\omega+(1-\omega)\frac{k+1}{t+1}\right)P(k),
\end{align}
\end{widetext}
which is valid for $k\geq 1$, while for $k=0$ only the loss term is present. Using the ansatz 
\begin{equation}
P(k,t) = \phi(k,t)e^{-\omega t}(1+t)^{-(1-\omega)(k+1)},
\end{equation}
where the time dependence of the degree distribution is again made explicit, the rate equation can be cast into (differential) recursion relations for the unknown functions $\phi(k,t)$:
\begin{equation}
\frac{\partial}{\partial t}\phi(k,t)=(1+t)^{1-\omega}\left(\omega+(1-\omega)\frac{k}{1+t}\right)\phi(k-1,t).
\end{equation}
Solving these equations with the correct initial condition, $\phi(k,t=0) = \delta_{k,0}$, and boundary condition $\phi(-1,t)=0$, yields
\begin{align*}
P(0,t)&=(1+t)^{-(1-\omega)}e^{-\omega t},\\
P(1,t)&=e^{-\omega t}\left(\frac{\omega}{2-\omega}(1+t)^\omega+(1+t)^{-(1-\omega)}\right.\\
&\left.\quad\quad-\frac{2}{2-\omega}(1+t)^{-2(1-\omega)}\right),
\end{align*}
for the lowest degrees. For general $k$, we can write: 
\begin{equation}
P(k,t) = g(k,1+t)e^{-\omega t},
\end{equation}
where the highest power of $1+t$ in $g(k,1+t)$ equals $k+\omega-1$. 
After an initial transient regime, the degree distribution thus follows a Poisson-like behaviour $P(k,t) \propto (1+t)^{k+\omega-1}e^{-\omega t}$. For smaller times, the details of $g(k,t)$ are important, but in none of the regimes a scale-free network is obtained.

\subsubsection{Approximation for large $\alpha$}
Before we consider the case of large $\alpha$, let us describe the limit $\alpha=+\infty$. The growth rule then selects (two of) the nodes with the smallest degrees and connects them. Consequently, only two node degrees will have a non-zero density and they will evolve linearly in time. For instance, the degree distribution in the pair model for small times ($0<t<1$) is:
\begin{align}\nonumber
P^p(k=1,t)&=1-t\quad\text{ and  }\quad P^p(k=2,t)=t.
\end{align}
Moreover, both the bachelor and the \pairmodel have exactly the same evolution for $\avk\geq 1$. Indeed, in the early stages ($t\leq 1$) of the bachelor model, all nodes are combined into pairs, which yields exactly the initial structure of the pair model when $\avk = 1$. In the following, we focus on the degree distribution in the initial stages ($0<t<1$, i.e., $1\leq\avk\leq 2$) of the pair model for large but finite $\alpha$. Note that other times can be considered with similar arguments.

In the initial stages, most nodes have degree one or two, but, for large but finite $\alpha$, a significant fraction of the nodes has three links. The probability that these are formed is proportional to $\ef{2}^p=2^{-\alpha}$. In the following this constant will be called $\epsilon \equiv 2^{-\alpha}$. We study the asymptotic case using an expansion of the constants $A(q,k)$, the function $x(t)$ and the differential equation \eqref{dvglx} up to first order in $\epsilon$. Straightforward calculations yield
\begin{subequations}\label{degree_alpha_infin}
\begin{align}
P^p(k=1,t)&=1-t - \twee [t+\ln(1-t)],\\
P^p(k=2,t)&=t+2 \twee[t+\ln(1-t)],\\
P^p(k=3,t)&=-\twee[t+\ln(1-t)].
\end{align}
\end{subequations}
Similar expressions for the \bachelormodel are obtained using the transformation of \eqref{n_pair_single}. 
 The expansion for the degree distribution is valid as long as  $|\twee \ln(1-t)| \ll 1$ and $(3/2)^{-\alpha}\ll 1$ \cite{phdthesis}. The approximation therefore breaks down if $t$ approaches one, a limitation that stems from the initial assumption $1 < \avk < 2$. If $\avk > 2$, the nodes will most likely have degree two and three, with a growing fraction of nodes with four links. The assumptions also break down if $\alpha$ is too small which is related to the neglect of factors $n^{-\alpha}$ for $n>2$. Note that in case either one of these conditions is not met, the approximation can be improved by the addition of the higher-order terms $n^{-\alpha}$ for $n>2$.

\begin{figure}[t]
\centering
\includegraphics[angle=0,width = .5 \textwidth]{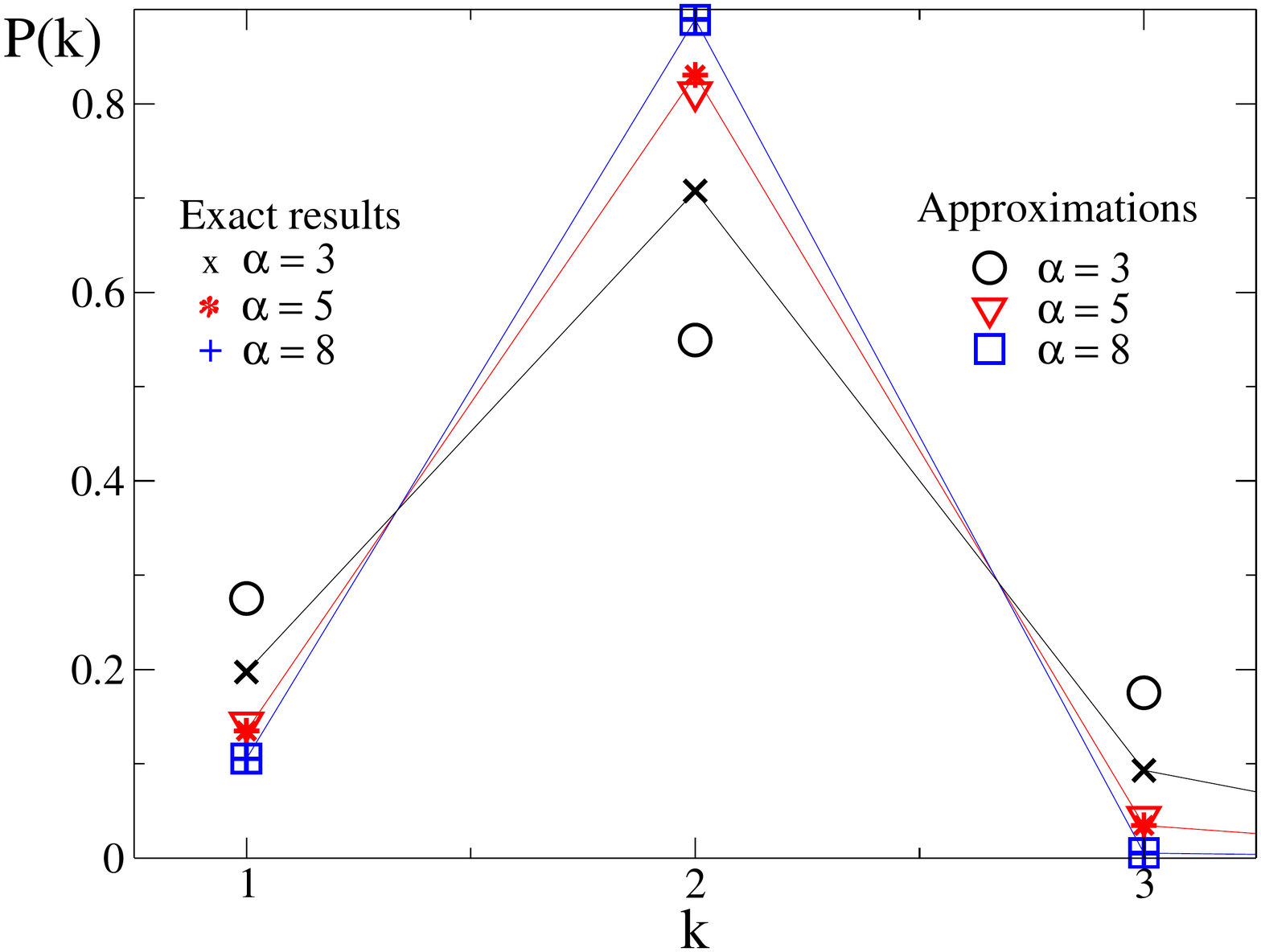}
\caption[Approximations for the degree distribution for large $\alpha$]{(Colour online) Comparison of the approximate degree distribution, \eqref{degree_alpha_infin}, and the exact degree distribution in networks with $\avk = 1.9$ for the \pairmodelpunt. Approximate results are shown as circles ($\alpha=3$), triangles ($\alpha = 5$) and squares ($\alpha = 8$). The exact results, which are connected with lines, are obtained by solving \eqref{ansatz1}, \eqref{A_coeff} and \eqref{determinex1} and are indicated with crosses ($\alpha = 3$), stars ($\alpha = 5$) and plus-signs ($\alpha = 8$). \label{figuurexpansiedistributie}}
\end{figure}

In Fig.~\ref{figuurexpansiedistributie} we compare the approximate degree distribution of \eqref{degree_alpha_infin}  with the exact results for networks at time $t=0.9$ ($\avk = 1.9$) for different values of $\alpha$. The exact results are derived from the rate equations and are verified using simulations. The values of the test parameters, $(|\twee \ln(1-t)|$ and $(3/2)^{-\alpha})$ are ($0.288$, $0.296$) for $\alpha=3$, ($0.072$, $0.131$) for $\alpha=5$ and finally ($9\cdot 10^{-3}$, $3.9\cdot 10^{-2}$) for $\alpha=8$. Clearly our approximation is poor for $\alpha = 3$ and improves upon increase of $\alpha$. The graph shows that the approximation overestimates the number of nodes with degrees one and three, while the number of nodes with degree two is underestimated. These effects are caused by a prominent (negative) feedback mechanism in the differential equation for $P(1)$. For large $\alpha$, good agreement between the exact and the approximate degree distributions is found. For instance, for $\alpha = 8$, the deviations are always smaller than one percent.

 In sum, we determined the degree distribution in the bachelor and the pair models. For general $\alpha$, we have to rely on numerical techniques, but analytical results can be found for some special cases. Although it may appear so far that the relation between the bachelor and the pair models is trivial, in the next section it will become clear that distinctions are present when looking at the degree correlations.

\section{Degree correlations\label{sectiondegreedegreecorrelations}}
General degree-dependent linking probabilities cause correlations in the constructed networks. Here, we study the \textit{degree correlations}, which describe relationships between nodes because of their degrees.  The nodes may be nearest neighbours but we also introduce correlations between nodes which are farther apart. Similar correlations are also present in scale-free networks created by the Barab\'{a}si-Albert linking process, as was proven by Barrat and Pastor-Satorras using a rate equation approach  \cite{barrat2005}.
In the first subsection below, we study the evolution laws obeyed by nearest-neighbour and next-nearest-neighbour-degree correlations during general network growth processes with a fixed number of nodes and degree-dependent link addition probabilities. We subsequently apply the formalism to our two diversifications.

\subsection{Nearest-neighbour degree correlations}
Nearest-neighbour degree correlations are usually characterised using the conditional \nearne probability $P_n(k|q)$, which expresses the probability that a nearest neighbour of a node with degree $q$ has degree $k$ \cite{dorogovtsev2007}. For notational simplicity, we introduce the related quantity $n_{k,q}$ defined as
\begin{equation}\label{Nkq}
n_{k,q}\equiv qP(q)P_n(k|q).
\end{equation}
The advantage of this description lies in the interpretation of $n_{k,q}$: one can prove that $Nn_{k,q}$ is the number of links between nodes with degrees $k$ and $q$ if $k\neq q$, while $Nn_{k,k}$ represents  twice the number of links between two nodes with degree $k$.
In simulations, an easily accessible quantity is the mean nearest-neighbour degree of a node with degree $q$, $\avknn_q$. The quantity $\avknn_q$ is related to $n_{k,q}$ by the expression
\begin{equation}\label{nearest_n}
\avknn_q = \frac{1}{qP(q)}\sum_k k\:n_{k,q}.
\end{equation}
In an uncorrelated network, the nearest-neighbour conditional probability $P_n(k|q)$ reduces to the uncorrelated nearest-neighbour distribution $P_n(k)$, defined as \cite{dorogovtsev2007}
\begin{equation}\label{onweer}
P_n(k) \equiv \frac{kP(k)}{\avk},
\end{equation}
and hence also $n_{k,q}$ factorises as $n_{k,q}=kqP(k)P(q)/\avk$. Moreover, the mean degree of a nearest-neighbour reduces to \cite{dorogovtsev2007}
\begin{align}
\avknn_q = \frac{\average{2}}{\avk}.
\end{align}
When $\avknn_q$ is plotted as a function of $q$, deviations from a constant behaviour thus advert to degree correlations. Note that two qualitatively different types of behaviour may be distinguished: assortative and disassortative mixing. The former occurs when hubs are preferentially linked to one another, while disassortative behaviour is present if sparsely connected nodes prefer to be linked with hubs \cite{newman2003a}.

The time evolution of $n_{k,q}$ is, in a mean-field approach, described by a rate equation. Simple, intuitive arguments lead to
\begin{align}\label{rateeqnkq}
\frac{dn_{k,q}}{dt} &= p_{k-1}n_{k-1,q}+p_{q-1}n_{k,q-1} \\& -n_{k,q}\left(p_k + p_q\right)+p_{k-1}p_{q-1}P(k-1)P(q-1), \nonumber
\end{align}
which is valid for $k,q > \kmin$. For $k = \kmin$, this rate equation only makes sense if we assume $n_{k,\kmin-1}= 0$ for all $k$. The first two terms on the right-hand-side of \eqref{rateeqnkq} describe the gain in $n_{k,q}$ if a new link  is attached to a node with degree $k-1$ or a node with degree $q-1$, which happens with probability $p_{k-1}P(k-1)$ and $p_{q-1}P(q-1)$, respectively. If a node with degree $k-1$ is chosen, the factor $N_{k,q}$ increases with the average number of nearest neighbours with degree $q$ of a node with degree $k-1$. This quantity is $(k-1)P_n(q|k-1)$, since each of the $k-1$ links of the chosen node has a probability $P_n(q|k-1)$ to lead to a node with degree $q$. Using \eqref{Nkq}, the gain term related to the selection of a node with degree $k-1$ is found to be $p_{k-1}n_{k-1,q}$, which is the first term in \eqref{rateeqnkq}. Similarly, the third term represents the loss in case a node with degree $k$ or $q$ is selected. The last term is special in the sense that it quantifies the possibility to add a new link between a node with degree $k-1$ and a node with degree $q-1$, in which case $N_{k,q}$ increases by one. A full and detailed derivation can be found in  Appendix~\ref{appendixrateeqnn}. Note that the rate equations can only be solved if the correct initial conditions are provided. For instance, for our bachelor model, in which no links are present at $t=0$, the initial condition is trivial:
\begin{equation}
n^b_{k,q}(t=0) = 0 \quad\quad \forall k,q.
\end{equation}
For the pair model, on the other hand, we need to express the presence of the initial links, which all run between nodes with degree one. Consequently,
\begin{subequations}\label{initialpair}
\begin{align}
n^p_{k,q}(t=0) &= 0 \quad\quad\textrm{if } (k,q)\neq(1,1),\\
n^p_{1,1}(t=0) &= 1,
\end{align}
\end{subequations}
as can be deduced from the interpretation of $Nn_{k,q}$.

Note that uncorrelated networks can only occur if the network growth rules allow them. If the uncorrelated expression for $n_{k,q}$ is inserted into the rate equation, \eqref{rateeqnkq}, a condition on the linking probabilities is found:
\begin{equation}\label{crit_no_corr}
p_{k-1}=\frac{kP(k)}{\avk P(k-1)}\quad \forall k>\kmin.
\end{equation}
Only linking processes which satisfy this condition with a constant-in-time right-hand-side, grow uncorrelated networks. We will explore this condition for the bachelor and the \pairmodel further on.

\subsection{Next-nearest-neighbour-degree correlations}
Degree-dependent attachment rules induce not only nearest-neighbour degree correlations, but also correlations amongst nodes which are farther apart. Here we extend our previous discussion and analysis to next-nearest-neighbour correlations, i.e., to correlations between nodes which are separated by two links. Analogously to the arguments in the previous subsection, it is possible to deduce a rate equation. We introduce $N_{q,k,s}$, the number of connected triplets in the network in which the middle node has degree $k$, while the other two nodes have degrees $s$ and $q$, and its fraction $n_{q,k,s} = N_{q,k,s}/N$. Note that $n_{s,k,q}=n_{q,k,s}\neq n_{k,s,q}$. 
In order to compare the theory with simulations, we introduce the mean degree of a node which is two links away from a node with connectivity $q$, $\langle k_{nn}\rangle_q$. This mean next-nearest-neighbour degree is related to $n_{s,k,q}$ by the relation
\begin{equation}\label{next_near}
\langle k_{nn}\rangle_q =\frac{ \sum_{s,k}s \: n_{s,k,q}}{\sum_{t,v}n_{t,v,q}}.
\end{equation}
 In an uncorrelated network, this equation reduces to
\begin{equation}
\langle k_{nn}\rangle_q = \frac{\avk^2}{\avk},
\end{equation}
which expresses, as expected, that the mean degree of a next-nearest neighbour equals the mean degree of a nearest neighbour.

A rate equation for $n_{k,q,s}$ can be found using arguments similar to those in the previous subsection:
\begin{widetext}
\begin{align}\label{rateeq2}
\frac{dn_{s,k,q}}{dt}&= -\left(p_k + p_q + p_s\right)n_{s,k,q}+ p_{s-1}n_{s-1,k,q}+p_{k-1}n_{s,k-1,q}+p_{q-1}n_{s,k,q-1}\nonumber\\&+p_{k-1}
p_{s-1}n_{k-1,q}P(s-1)+p_{q-1}p_{k-1}n_{k-1,s}P(q-1).
\end{align}
\end{widetext}
This equation is valid if $s$, $k$ and $q$ are larger than $\kmin$.
The first term on the right-hand side stems from the loss of $(s,k,q)-$triplets if a node with degree $s$, $k$ or $q$ is selected; analogously, the second and third terms express the gain if a node with degree $s-1$, $k-1$ or $q-1$ is selected. Finally, the last two terms indicate the gain in triplets if a link between nodes with degrees $s-1$ and $k-1$ or between nodes with degrees $k-1$ and $q-1$ is laid. Due to the random and local nature of the linking scheme no triangles are formed in the thermodynamic limit, or equivalently, the clustering coefficient vanishes for large networks as was verified using simulations. Note that with an analogous scheme, a rate equation for correlations of even higher order can be introduced. However, the computational complexity increases as the distance between the nodes increases and therefore we will not pursue such an endeavour here.

In the following, we apply the rate equations for the degree correlations to the pair and the bachelor model separately. We will also observe next-next-nearest-neighbour correlations in simulations using the mean degree of a node which is separated by three links from a node with connectivity $q$,
 $\langle k_{nnn}\rangle_q$. 

\subsection{Application to the \bachelormodel}
\subsubsection{Nearest-neighbour correlations}
\begin{figure}[t]
\centering
\includegraphics[angle=0,width = .5 \textwidth]{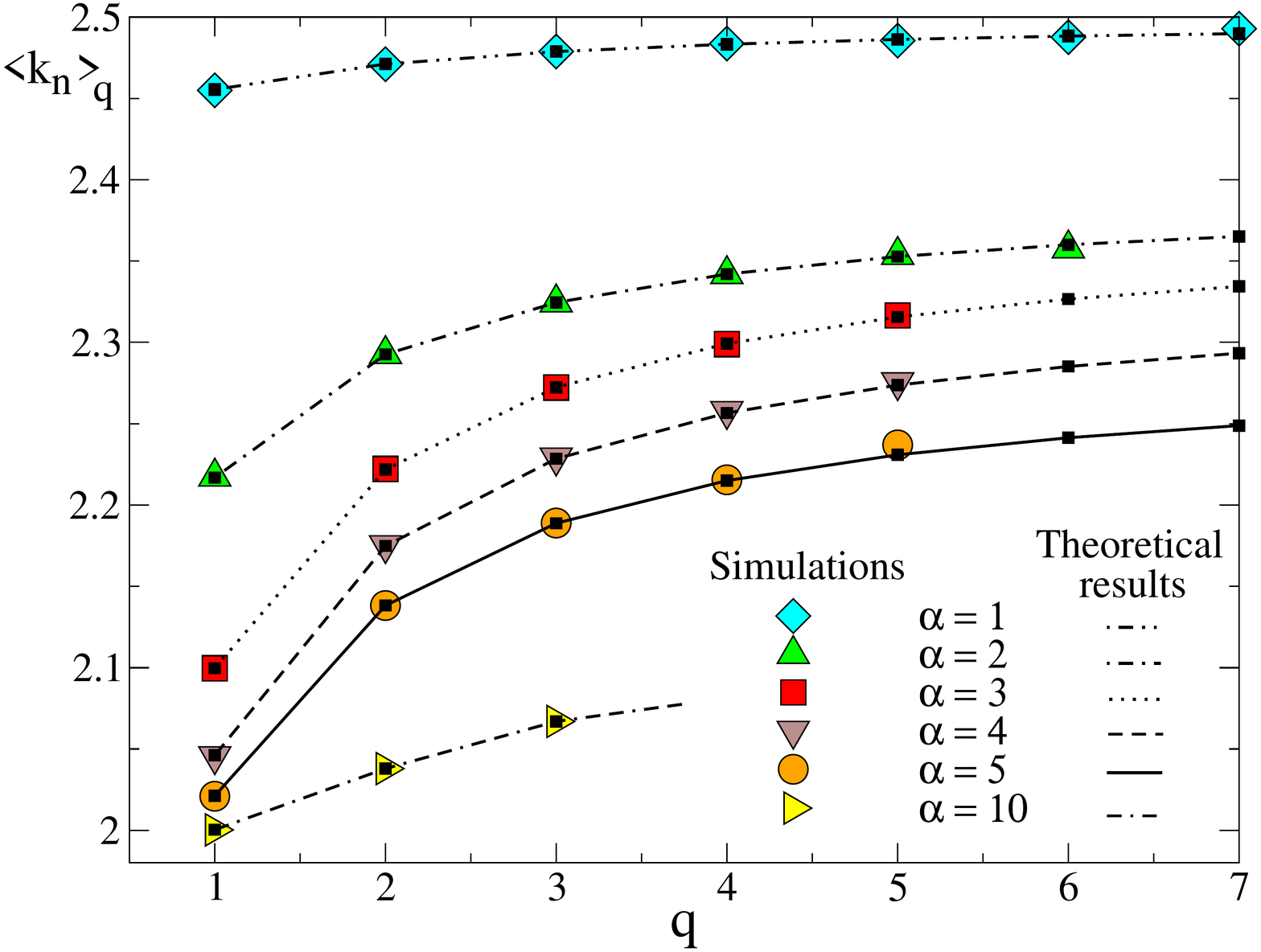}
\caption[Nearest-neighbour correlations in the \bachelormodelpunt.]{(Colour online) Nearest-neighbour correlations in the \bachelormodel for networks with $\avk = 2$. The figure shows the mean nearest-neighbour degree $\avknn_q$ as a function of the degree $q$ of a node. Both simulation data (symbols) and numerical solutions using the rate equation, \eqref{rateeqnkq}, (lines) are plotted. Results are shown for $\alpha = 1$ (blue diamonds, dot-dot-dashed line), $\alpha = 2$ (green up triangles, dot-dashed line), $\alpha = 3$ (red squares, dotted line), $\alpha = 4$ (brown down triangles, dashed line), $\alpha = 5$ (orange circles, full line) and $\alpha = 10$ (yellow right triangles, dot-dash-dashed line). All simulation data are averages over 100 realisations of the growth of a network with $10^6$ nodes. The standard errors are smaller than the symbol sizes. 
 \label{correlfig}}
\end{figure}

\begin{figure}[t]
\centering
\includegraphics[angle=0,width = .45 \textwidth]{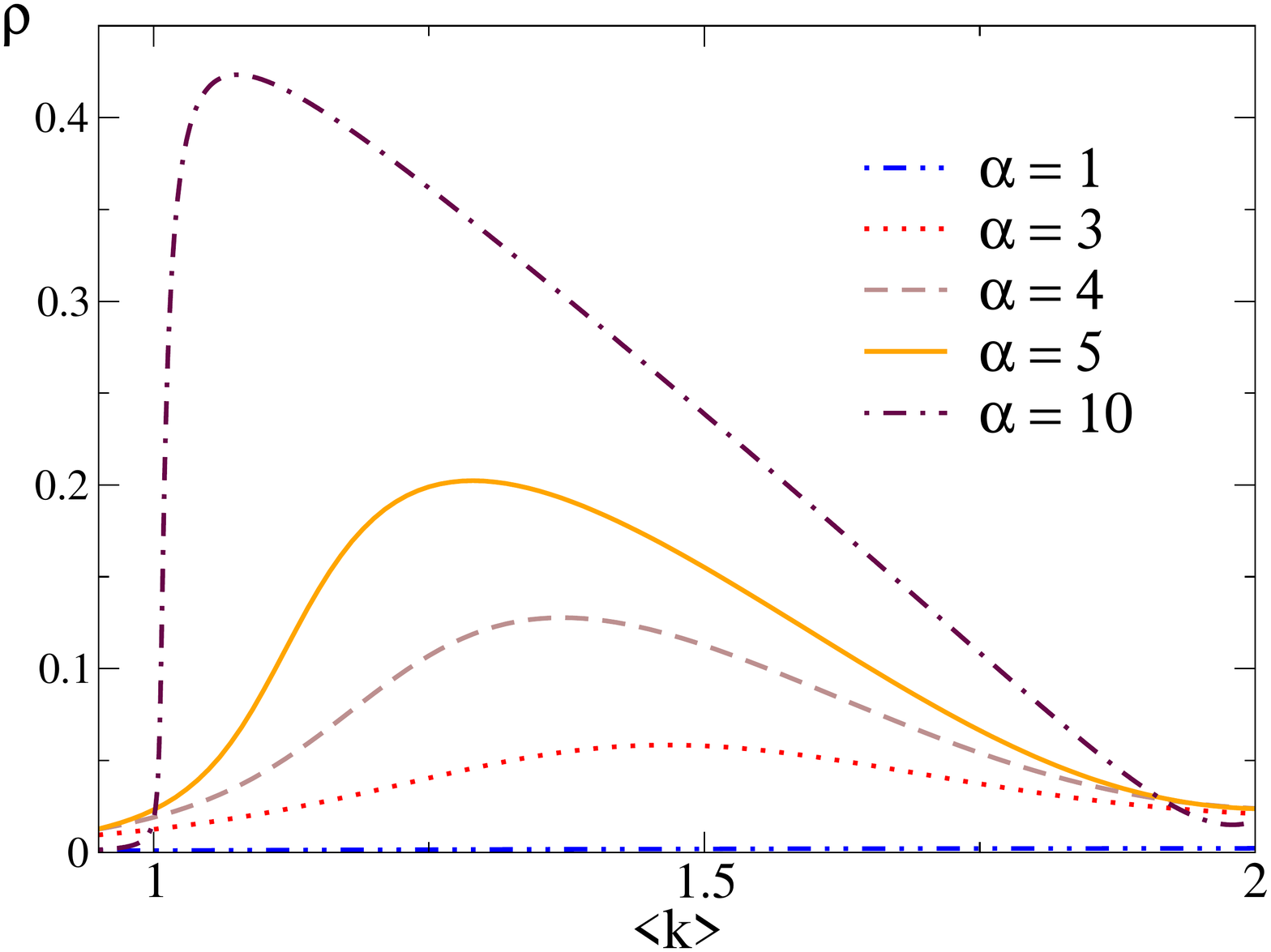}
\caption[Nearest-neighbour Pearson degree correlation $\rho$ in the \bachelormodelpunt.]{ (Colour online) Assortativity coefficient $\rho$ as defined in~\eqref{defrho} for the \bachelormodel as a function of the mean connectivity $\avk$. Results are obtained using numerical integration and are shown for $\alpha = 1$ (blue dot-dot-dashed line), $\alpha = 3$ (red dotted line), $\alpha = 4$ (brown dashed line), $\alpha = 5$ (orange full line) and $\alpha = 10$ (black dot-dashed line).
 \label{PearsonCoef}}
\end{figure}

In this section we study in detail the degree correlation caused by the degree-dependent attachment rule  in the bachelor model. In Fig.~\ref{correlfig}, simulation data for the mean nearest-neighbour degree in networks with $\avk =2$ are compared with numerical solutions to the rate equation, \eqref{rateeqnkq}, for some chosen values of  $\alpha$. The theoretical results clearly agree with the simulation data within the error margins. The rate-equation approach thus provides an accurate description of the nearest-neighbour degree correlations. In the remainder of this section, we will therefore fully exploit the strength of the rate equation approach to study the correlations in the bachelor model.

Although the network is correlated in general, for some values of $\alpha$ an uncorrelated network is retrieved. For $\alpha = 0$ and $\alpha = -1$, the degree distributions are exactly known (see \eqref{dda0} and \eqref{dda-1}) and some straightforward calculations yield that the attachment rule satisfies \eqref{crit_no_corr} in both cases. The absence of correlations for these special cases is also confirmed in simulations and, for $\alpha = 0$, it is a well-known property of Erd\H{o}s-R\'{e}nyi networks \cite{erdos1960,bollobas2001}. For $\alpha = -1$, on the other hand, it is a particularity of our model. In models with the same attachment rule but with a variable number of nodes, like the Barab\'{a}si-Albert model, the network is correlated due to the degree-dependent attachment rule \cite{krapivsky2000,barrat2004}. We conclude that these correlations can be attributed to the continual addition of nodes to the network.

By inserting the numerically obtained degree distribution into the rate equation for $n_{k,q}$ it is found that the network is assortatively mixed if $\alpha \notin \{-1,0\}$. Nodes with a large degree are thus preferentially linked to one another. However, the correlations are very small for negative $\alpha$. We quantify the importance of the correlations using the fractional difference
\begin{equation}\label{defdelta}
\mu_n \equiv  \left.\frac{|\avknn_{q=1} -\avknn_{q=5}|}{\avknn_{q=1}}\right|_{\avk = 2},
\end{equation}
which expresses the deviation of $\avknn_q$ from the constant curve observed in uncorrelated networks. As an alternative measure of degree correlation we use the assortativity coefficient $\rho$ as defined in Eq.~(2) in Ref.~\cite{newman2003b}:
\begin{align}\label{defrho}
\rho=\frac{\sum_q\left(\avk n_{q,q}-\left[qP(q)\right]^2\right)}{\avk^2-\sum_q\left[qP(q)\right]^2}.
\end{align}
Within the interval $[-1,0]$ for $\alpha$, the largest $\mu_n$ and $\rho$ occur around $\alpha \approx -0.65$, for which $\mu_n \approx 6\cdot 10^{-4}$ and $\rho=5\cdot 10^{-5}$, which are smaller than the error margins obtained in simulations. The correlations are thus negligible for negative~$\alpha$.
For positive $\alpha$, on the other hand, the nearest-neighbour correlations are much larger and will have a substantial influence on the percolation properties as will be discussed further on. For example for the case $\alpha = 1$, both $\mu_n$ and $\rho$ are two orders of magnitude larger than for $\alpha=-0.65$, as shown in Table 1. Fig.~\ref{correlfig} shows that the importance of the correlations for $\avk=2$ further increases with increasing $\alpha$ and attains a maximum around $\alpha \approx 4$. 
For values of $\alpha$ larger than four, the correlations diminish (see Table 1) and in the limit $\alpha\rightarrow \infty$ they become negligible. This behavior at $\avk=2$ can be also observed in Fig.~\ref{PearsonCoef} which shows $\rho$ as a function of $\avk$ for different values of $\alpha$. The network is fully assortative ($\rho=1$) in case all nodes connect to nodes of the same degree.  On the other hand, it can be verified that $\rho=0$ for an uncorrelated network. The assortativity coefficient $\rho$ is undefined when only one nonzero degree is present in the network. This is the case in the limit of infinite $\alpha$ when $\avk<1$ since then nodes have either degree zero or degree one. However, it is easily verified that in the limit $\alpha\rightarrow \infty$, $\rho$ converges to $\rho=0$. Therefore $\rho$ jumps to a nonzero value (here $1/2$) at $\avk=1$ since new links will introduce nodes of degree two that have one neighbour of degree one and one of degree two. This abrupt change near $\avk=1$ can be noticed in Fig.~\ref{PearsonCoef} for the case $\alpha=10$.


\begin{table}
\centering
\caption{Nearest-neighbour degree correlation coefficients $\mu_n$ and $\rho$ as defined in \eqref{defdelta} and \eqref{defrho}, respectively, against different values of $\alpha$ when $\avk=2$ for the \bachelormodelpunt.~\label{table1}}

\begin{tabular}
{||l||l|l||} 

\hline  
$\alpha$ & $\mu_n$ & $\rho$ \\ \hline\hline
-0.65&$6\cdot 10^{-4}$&$5\cdot 10^{-5}$\\ \hline 
0    & 0 & 0 \\ \hline
1    & $0.012$ & 0.0021\\ \hline
3    & 0.103 & 0.0210\\ \hline 
4    & 0.111 & 0.0238\\ \hline
5    & 0.107 & 0.0236\\ \hline
10   & 0.043 & 0.0172\\\hline
15   & 0.019 & 0.0114\\ \hline\hline
\end{tabular}
\end{table}

\subsubsection{Next-nearest-neighbour correlations}
\begin{figure}[t!]
\subfigure[Mean next-nearest-neighbour degree $\langle k_{nn}\rangle_q$.\label{correlNN}]{\includegraphics[angle=0,width = .5 \textwidth]{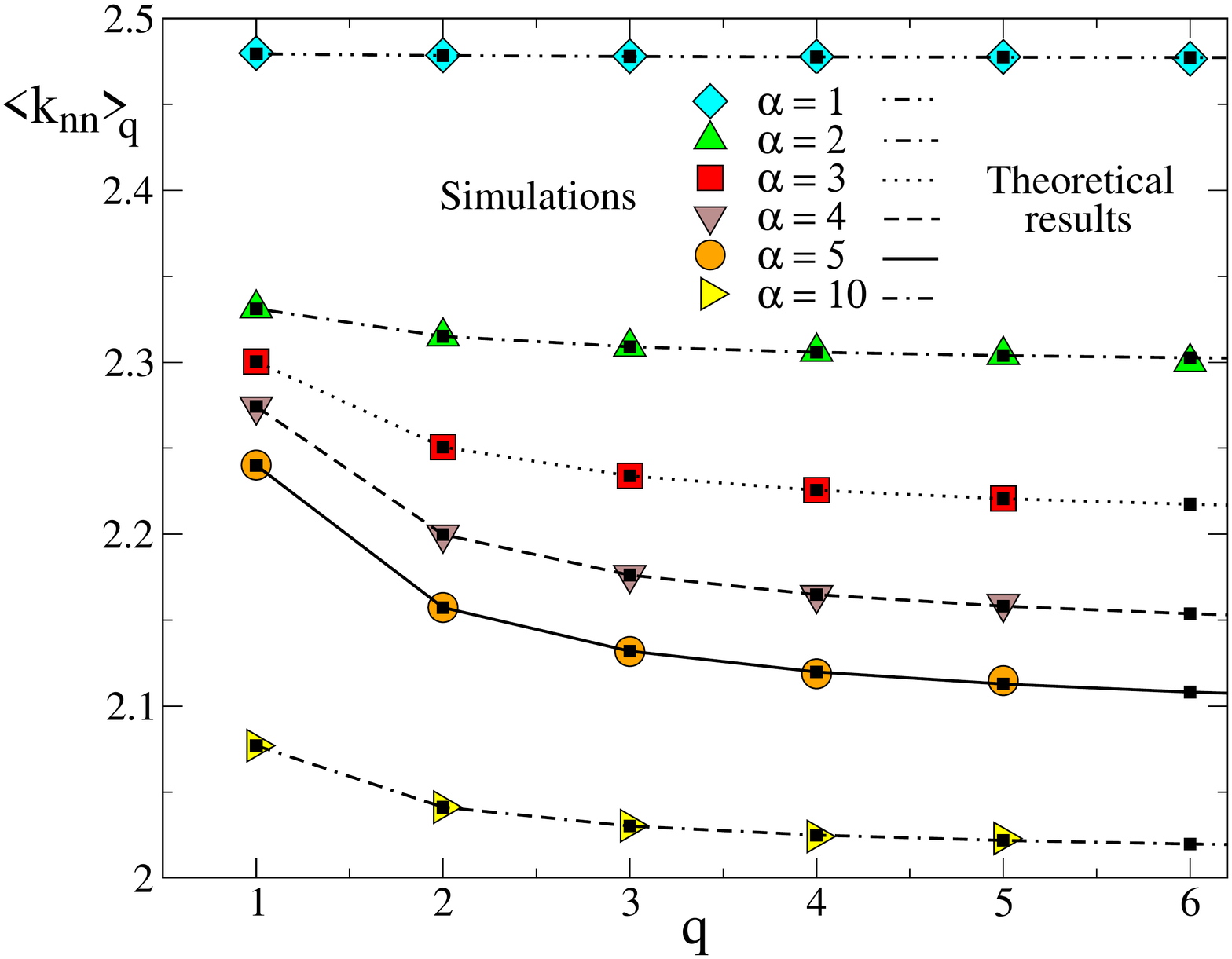}}
\subfigure[Mean next-next-nearest-neighbour degree $\langle k_{nnn}\rangle_q$.\label{correlNNN}]{\includegraphics[angle=0,width = .5 \textwidth]{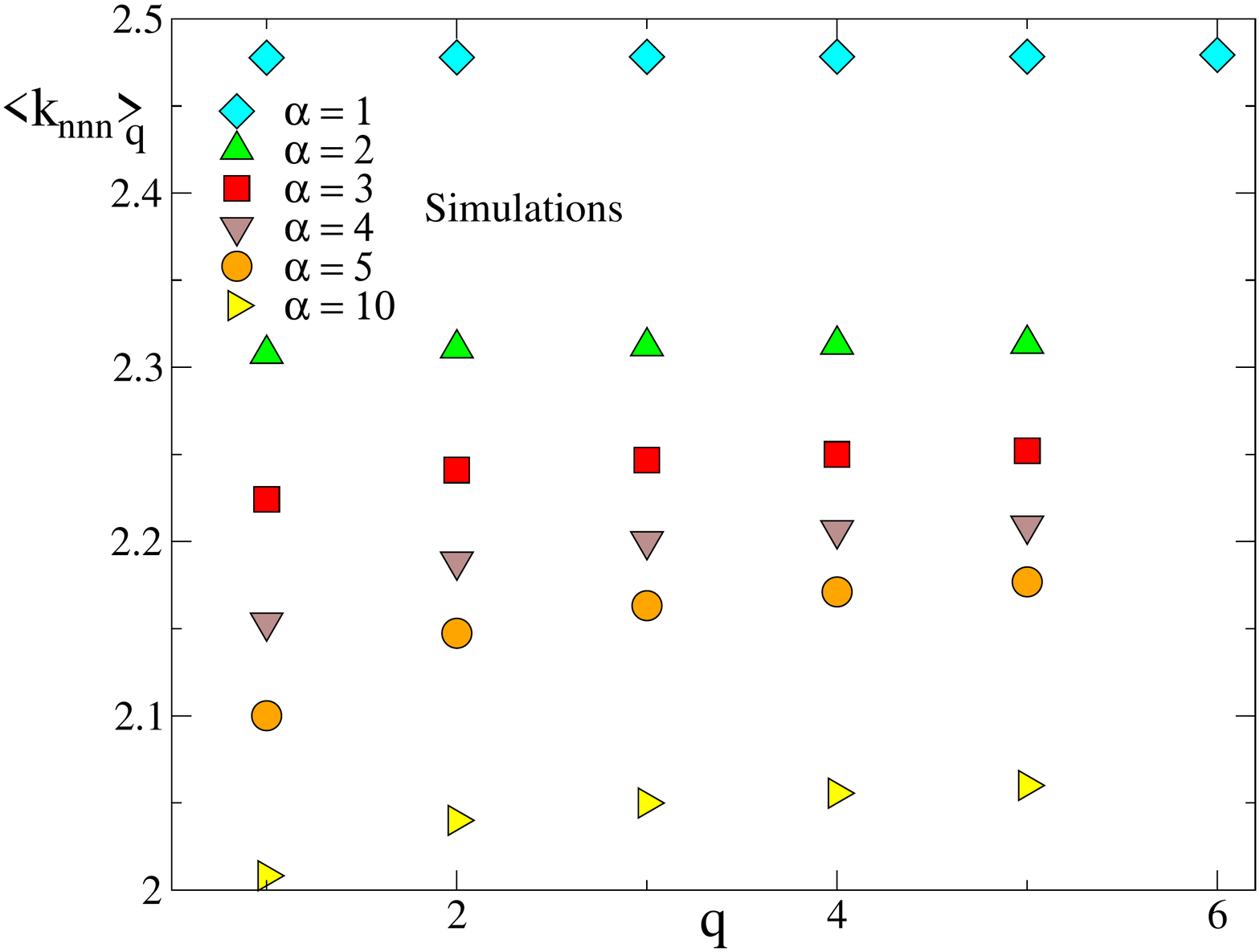}}
\caption{(Colour online) Higher-order degree correlations in the \bachelormodel for networks with $\avk = 2$,  as expressed by the mean next nearest-neighbour degree (Fig.~\ref{correlNN}) and the mean next-next-nearest-neighbour degree (Fig.~\ref{correlNNN}). Simulation data are shown with symbols and in the top figure we also show numerical solutions to the rate equation \eqref{rateeq2} using lines. The cases considered are $\alpha = 1$ (blue diamonds, dot-dot-dashed line), $\alpha = 2$ (green up triangles, dot-dashed line), $\alpha = 3$ (red squares, dotted line), $\alpha = 4$ (brown down triangles, dashed line), $\alpha = 5$ (orange circles, full line) and $\alpha = 10$ (yellow right triangles, dot-dash-dashed line). All simulation data are averages over 100 realisations of the growth of a network with $10^6$ nodes. The standard error is smaller than the symbol size. 
\label{dubbelefiguur}}
\end{figure}
Let us now focus on the next-nearest-neighbour correlations for the \bachelormodelpunt. The presence of non-zero correlations for positive $\alpha$ is conspicuous from Fig.~\ref{correlNN}, in which simulation data for  $\langle k_{nn}\rangle_q$ are shown for networks with $\avk = 2$. The figure also compares simulations with analytical results obtained by numerically  solving the rate equations \eqref{rateeqnkq} and \eqref{rateeq2} simultaneously. The theoretical results clearly match the simulations for all values of $\alpha$. Once again the rate equation approach correctly predicts the network properties.

Fig.~\ref{correlNNN} moreover shows the presence of next-next-nearest-neighbour-degree correlations. While the next-nearest neighbours clearly show disassortative mixing, the next-next-nearest neighbours are assortatively mixed. A link attached to a neighbour of a node with many links has thus a higher probability to lead to a low-degree node than to another high-degree node. A node reached after randomly following three links is most likely again a node with a more-than-average amount of links. For negative $\alpha$, simulations again indicate that the correlations are very small, while they are completely absent in the special cases $\alpha = 0$ and $\alpha =-1$.
For both  $\langle k_{nn}\rangle_q$ and $\langle k_{nnn}\rangle_q$, the importance of the correlations anew attains a maximum around $\alpha=4$. Furthermore, the effect of the correlations decreases as the distance between the nodes gets larger. These effects are quantified by the fractional differences $\mu_{nn}$ and $\mu_{nnn}$, which are defined analogously to $\mu_n$ (see \eqref{defdelta}) but in terms of  $\langle k_{nn}\rangle_q$ and $\langle k_{nnn}\rangle_q$ instead of $\langle k_{n}\rangle_q$. For instance,
\begin{equation}\label{defdeltann}
\mu_{nn} \equiv  \left.\frac{|\langle k_{nn}\rangle_{q=1} -\langle k_{nn}\rangle_{q=5}|}{\langle k_{nn}\rangle_{q=1}}\right|_{\avk = 2}.
\end{equation}
Simulations yield $\mu_n = 0.107$, $\mu_{nn} = 0.056$ and $\mu_{nnn} = 0.037$ for $\alpha = 5$, while $\mu_n = 0.043$, $\mu_{nn} = 0.027$ and $\mu_{nnn} = 0.023$ for $\alpha = 10$. The correlations between nearest neighbours are thus more important than the correlations between next-next-nearest neighbours. For $\alpha = 5$, the importance of the correlations falls off by about a factor of three from $\mu_n$ to $\mu_{nnn}$. For $\alpha = 10$, the reduction is about a factor of two.

We conclude that, in the bachelor model, degree correlations are present if $\alpha \notin \{-1,0\}$. Although the correlations are negligible for negative $\alpha$, they are significant for positive $\alpha$, as we will indicate when discussing the percolation properties. In all regimes, the correlations are well described by analytical results obtained using a rate equation approach.

\subsection{Application to the \pairmodel}\label{sect_pair1}
The description of correlations in the \pairmodel is more involved than that in the \bachelormodelpunt. The initial-time links which, at time zero, combine all nodes into pairs, introduce initial degree correlations. Furthermore, additional correlations arise as a consequence of the degree-dependent attachment rule. The evolution of the total correlations is anew described by the rate equations, \eqref{rateeqnkq} and \eqref{rateeq2}, if the correct initial conditions, \eqref{initialpair}, are applied. We will, however, not solve these relations numerically, but instead proceed analytically by disentangling the initial-link and finite-time correlations.

\subsubsection{Disentanglement of initial-link and finite-time correlations}
To disentangle the initial-link  and the finite-time correlations, we write $n_{k,q}$ as the sum
\begin{equation}\label{sum_links}
n_{k,q}^p=n_{k,q}^i+n_{k,q}^f,
\end{equation} 
where the first term represents the initial-time links and the second term the links which are laid later on. 

Consider first the initial-time links between nodes of degree $k$ and $q$. Since these links were laid randomly and each node has exactly one such link, connected nodes are initially uncorrelated and will remain so at finite time such that the fraction $n_{k,q}^{i}$ is at all times:
\begin{equation}\label{pair_corr}
n_{k,q}^i= P^p(k) P^p(q).
\end{equation} 
We continue by a substitution of \eqref{sum_links} and \eqref{pair_corr} into the dynamical equation, \eqref{rateeqnkq}. It is then straightforwardly obtained that the dynamical equation for $n_{k,q}^f$ is exactly the one of \eqref{rateeqnkq} but with $n_{k,q}$ replaced by $n_{k,q}^f$. Since the initial condition $n_{k,q}^f(t=0)=0$ coincides with that for the total correlations of the \bachelormodel and $f^b_k=f^p_{k+1}$, it is straightforwardly derived that 
\begin{equation}\label{nfkq}
n_{k,q}^f=n_{k-1,q-1}^b.
\end{equation} 
In other words, the nearest-neighbour degree correlations of the \pairmodel can be decomposed into initial-link correlations, which satisfy  \eqref{pair_corr}, and correlations which, after a shift of degrees, are exactly equal to those appearing in the \bachelormodelpunt.

In order to further disentangle the initial-link  and the finite-time correlations, we make a crude approximation by neglecting the latter. In \eqref{nfkq}, this implies a replacement of $n^b_{k-1,q-1}$ by its uncorrelated counterpart, which yields
\begin{equation}\label{nofinitetimecorrel}
n^f_{k,q}=(k-1)(q-1)\frac{P^p(k)P^p(q)}{t}.
\end{equation} 
Insertion in \eqref{sum_links} yields the normalised expression for $n_{k,q}^p$:
\begin{equation}\label{pair_corr_unc}
n_{k,q}^p= P^p(k)P^p(q) +(k-1)(q-1)\frac{P^p(k)P^p(q)}{t}.
\end{equation} 
Straightforward calculations then provide an approximate expression for the mean nearest-neighbour degree:
\begin{equation}\label{pkqinitial}
\avknn_q = -\frac{1}{q}\left(\frac{\average{2}-(\avk^p)^2}{\avk^p-1}\right)+\frac{\average{2}-\avk}{1+\avk^p}.
\end{equation}
These degree correlations, solely caused by the initial-time links, are clearly assortative since $\avknn_q$ increases with increasing degree $q$. Note that the parameter $\alpha$ appears only implicitly in the expression for $\avknn_q$ through the second moment of the degree distribution. 
In the following we discuss the correlations described by \eqref{pkqinitial} and compare this approximate relation with simulation data. 

\subsubsection{Results}
\begin{figure}[t]
\centering
\includegraphics[angle=0,width = .5 \textwidth]{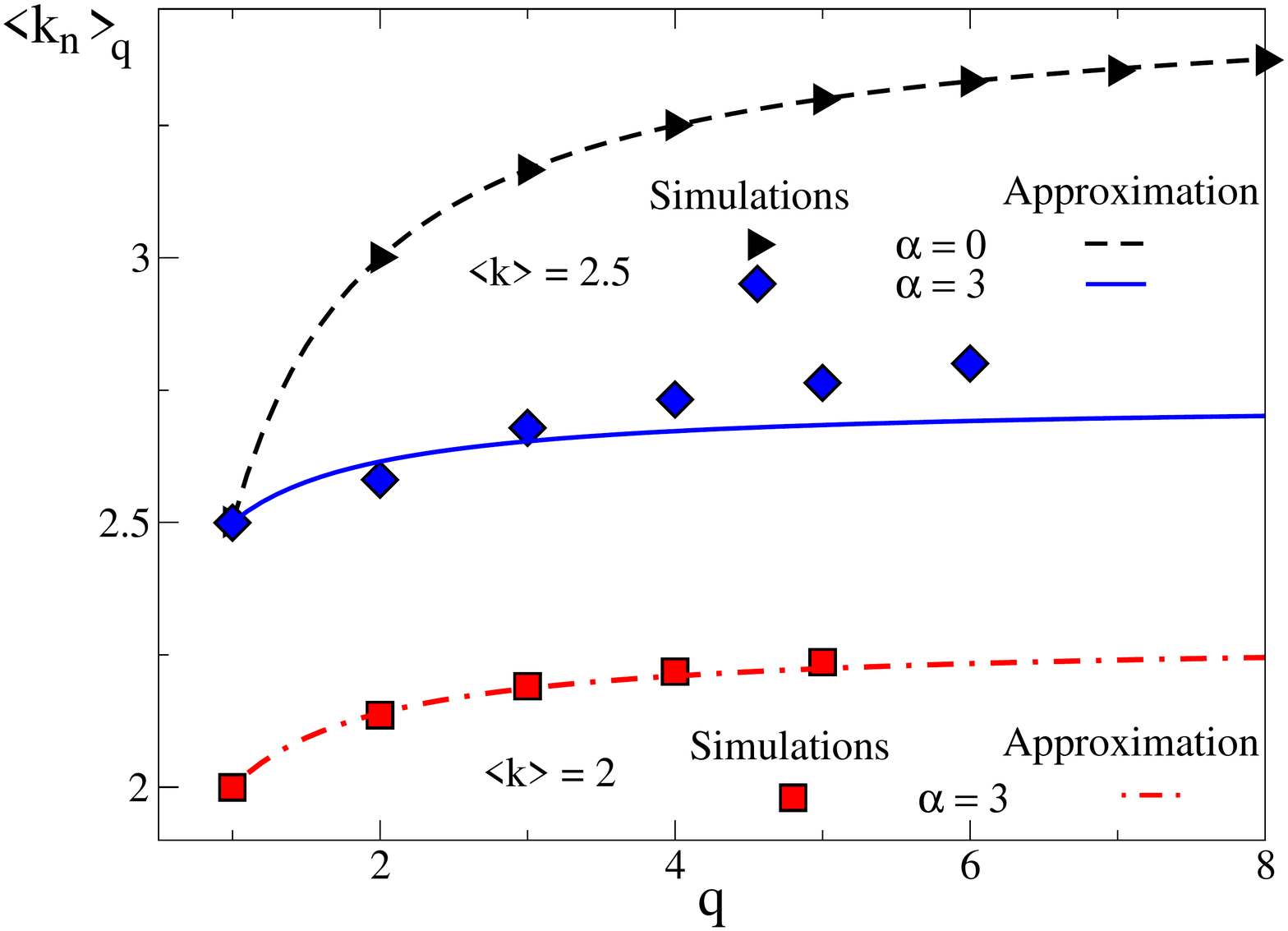}
\caption[Nearest-neighbour degree correlations in the pair model.]{(Colour online) Degree correlations as expressed by the mean nearest-neighbour degree $\avknn_q$ (\eqref{nearest_n}) against node degree $q$ in the \pairmodelpunt. The figure shows both simulation data (symbols) and the results of disregarding finite-time correlations, \eqref{pkqinitial} (lines), for two different time-steps. For $\avk = 2.5$ we show the cases $\alpha = 0$ (black triangles and dashed line)  and $\alpha =3$  (blue diamonds and full line), while for $\avk = 2$ we only show results for $\alpha =3$ (red squares and dot-dashed line). All simulation data are averages over 100 realisations of the growth of a network with $10^6$ nodes. The standard errors are smaller than the symbol sizes. \label{figinitialpkq}}
\end{figure}
In Fig.~\ref{figinitialpkq}, we compare the mean nearest-neighbour degree obtained by disregarding finite-time correlations, \eqref{pkqinitial}, with simulation data for two different values of $\alpha$. For $\alpha = 3$, we give results for two different times: $\avk =2$ and $\avk = 2.5$. For the former time-step, the agreement between \eqref{pkqinitial} and the simulations is remarkably good. When $\avk = 2.5$, however, there is a significant discrepancy. Simulations indicate that this deviation grows in time. At an early stage in the linking process, the initial-link correlations thus dominate the correlations due to the degree-dependent attachment for the pair model and \eqref{pkqinitial} provides a good approximation for the nearest-neighbour correlations.  In the latter stages, the correlations due to the degree-dependent attachment are also important and \eqref{pkqinitial} becomes inaccurate. In this regime, the complete rate equation, \eqref{rateeqnkq}, should be solved numerically using the correct initial conditions. For the special case $\alpha = 0$, on the other hand, Fig.~\ref{figinitialpkq} indicates that the analytical approximation without finite-time correlations, \eqref{pkqinitial}, clearly matches the simulation data at the moment when $\avk = 2.5$. In this random linking limit, finite-time correlations are absent and the approach outlined in the previous section thus becomes exact. Note also that, in contrast to the behaviour for the bachelor model, the network is clearly strongly correlated even for random network growth.

The correlations are again quantified by the quantity $\mu_n$ defined in \eqref{defdelta}. Using \eqref{pkqinitial}, we obtain $\mu_n = 0.12$ for $\alpha = 3$, $\mu_n = 0.4$ for $\alpha =0$ and $\mu_n = 0.80$ for $\alpha = -1$. Within the limitations set  for $\alpha$ the largest correlations are thus observed for $\alpha = -1$, which anew contrasts sharply with the behaviour for the \bachelormodelpunt. Also higher-order correlations exist in the \pairmodel in agreement with the separation of the correlations in \eqref{sum_links}. Simulations and numerical solutions to the rate equations indicate that their properties are reminiscent of the \bachelormodelpunt: the next-nearest-neighbour correlations are disassortative while the correlations between nodes separated by three links are again assortative. The importance of the correlations again decreases as the nodes are farther apart.

We conclude that, although the degree distributions in both models are related in a trivial manner, the degree correlations turn out to be different. The correlations in the \pairmodel are much larger due to the presence of correlations at the start of the linking procedure. In the early stages of the linking procedure, these initial-link correlations dominate the correlations due to the preferential attachment and the latter can therefore be neglected.

\section{Percolation properties\label{sectionpercolation}}
We first briefly recall the main ideas of the percolation problem, which is concerned with the connectivity of the network under a link addition process \cite{stauffer1994}. At the stage of initial network growth, the amount of clusters in the network is proportional to the amount of nodes. As more and more links are introduced in the network, the separate clusters join into larger clusters. At the critical percolation point a phase transition to a network with one giant connected component occurs. The transition is described using the order parameter $\mathcal{S}$ which quantifies the fraction of nodes in the largest cluster of the network.  Below the critical threshold, the only solution is $\mathcal{S}=0$, which corresponds to a network without a giant cluster, while a solution with a non-zero $S$ exists above criticality, that is when $t>t_c$, or, equivalently, $\avk>\avk_c$. 

\subsection{Generating functions formalism}
Analytical results are obtained using the generating functions formalism, which is described in more detail in Ref.~\cite{newman2001}. First, we assume equivalence of all links in the networks as is valid for the \bachelormodelpunt. However, such theory does not apply to the \pairmodel since the initial-time links are not equivalent to the finite-time links.  A corrected formalism will be presented later on. 

In an \textit{uncorrelated} network, the size of the giant cluster $\mathcal{S}$ is given by the system of self-consistent equations
\begin{subequations}\label{percolbenad1_single}
\begin{align} 
u&=\sum_q P_n(q)u^{q-1},\\
 \mathcal{S} &= 1-\sum_k P(k) u^k,
\end{align}
\end{subequations}
where $P_n(q)$ is the uncorrelated nearest-neighbour distribution introduced in \eqref{onweer} and $u$ represents the probability that a randomly chosen link leads to a cluster of finite size \cite{newman2001}. If the evolution of the degree distribution is fully known, the system can be solved numerically using a simple iterative scheme. However, degree correlations may have a strong influence on the percolation features of the network~\cite{goltsev2003,vazquez2003,goltsev2008}. The self-consistent system is straightforwardly generalised to a network with degree correlations solely between nearest neighbours \cite{newman2001}:
\begin{subequations}\label{percolbenad2_single}
\begin{align} 
u_k&=\sum_q P_n(q|k)\left(u_q\right)^{q-1},\label{percola}\\
 \mathcal{S} &= 1-\sum_k P(k)\left(u_k\right)^k.\label{percolb}
\end{align}
\end{subequations}
Here $P_n(q|k)$ is the \nearne conditional probability and $u_k$ is the probability that, upon starting from a random node of connectivity $k$, one follows a random link to the other side and arrives at a cluster of finite size.  The system can again be solved numerically with a simple iterative scheme. The effect of the next-nearest-neighbour correlations can be captured using the numerical results for $n_{k,q,s}$. The size of the largest cluster is given by self-consistent equations for the nearest-neighbour order parameters  $u_{k}$ and the next-nearest-neighbour order parameters $u_{k,q}$:
\begin{subequations}\label{percolbenad3_single}
\begin{align}
u_{q,k} &= \sum_s P_{nn}(s|q,k)\left(u_{s,q}\right)^{s-1},\\
u_k &= \sum_q P_n(q|k)\left(u_{q,k}\right)^{q-1},\\
\mathcal{S} &= 1- \sum_k P(k) \left(u_k\right)^k,
\end{align}
\end{subequations}
where $u_k$ is defined as before and $u_{q,k}$ is its generalisation to next-nearest neighbours. The probability $P_{nn}(s|q,k)$ indicates the probability that a node with degree $q$ is connected to a node with degree $s$, given that the $q$-node was reached at the end of a link emerging from a node with degree $k$. Therefore:
\begin{equation}
P_{nn}(s|q,k) = \frac{n_{s,q,k}}{\sum_{t}n_{t,q,k}}.
\end{equation}
Here $n_{s,q,k}$ satisfies the rate equation, \eqref{rateeq2}. The proof of \eqref{percolbenad3_single} is a simple extension of the proof for the self-consistent set for networks with only nearest-neighbour correlations.

\subsection{Application to the \bachelormodel}
\begin{figure}[t]\centering
\includegraphics[angle=0,width = .5 \textwidth]{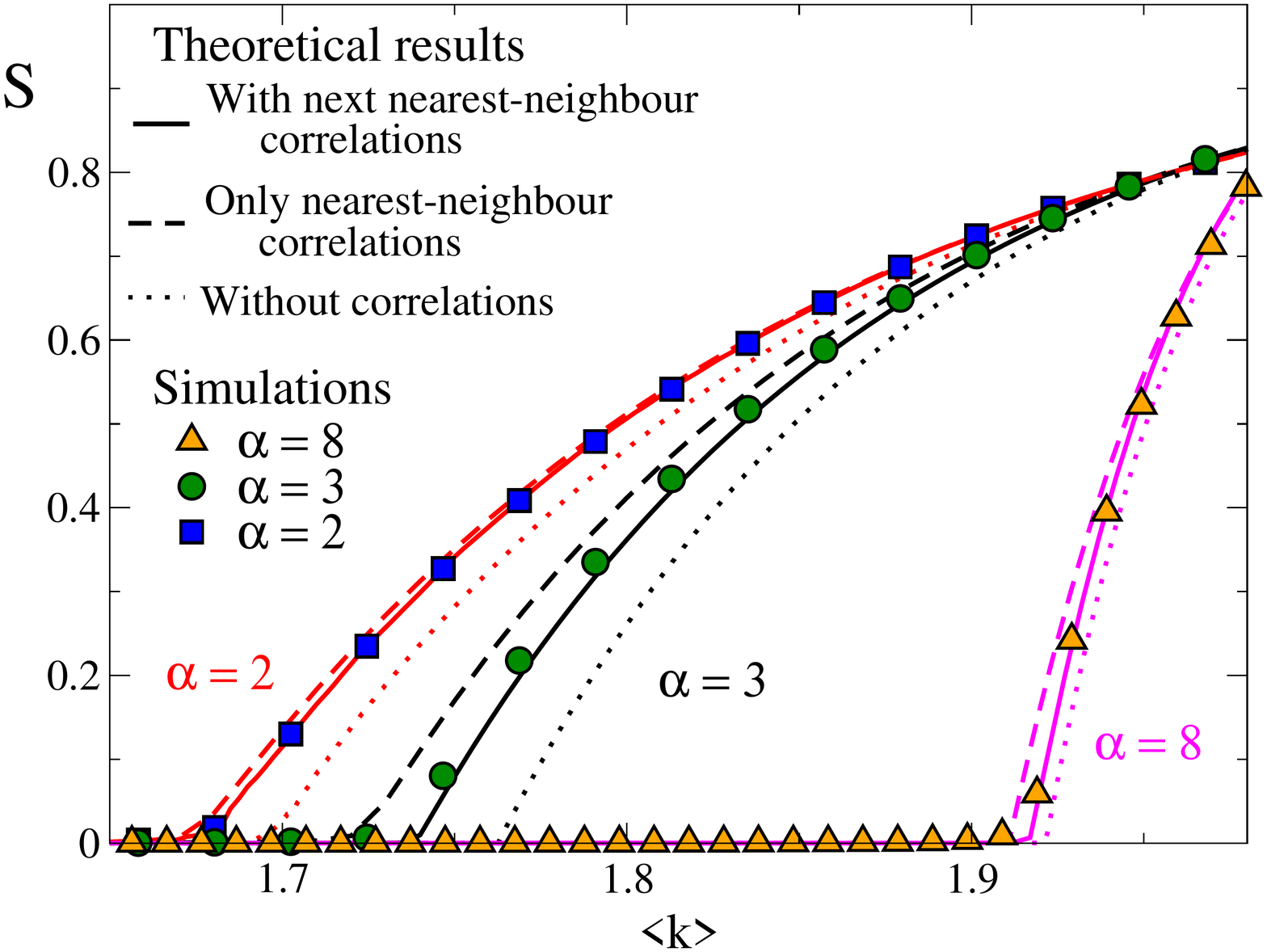}
\caption[The percolation transition in the \bachelormodelpunt.]{(Colour online) The percolation transition in the \bachelormodelpunt. The figure shows the size of the largest cluster as a function of the mean degree $\avk$ in the network. Both simulation data (symbols) and various theoretical approximations (lines) are shown for $\alpha = 2$ (red, upper three lines and blue squares), $\alpha =3$ (black, middle three lines and green circles) and $\alpha = 8$ (magenta, lower three lines and orange triangles). The dotted lines show the numerical solutions of the generating functions theory without correlations, \eqref{percolbenad1_single}, the dashed lines indicate the theoretical results with only nearest-neighbour correlations, \eqref{percolbenad2_single}, and finally the full lines use also next-nearest-neighbour correlations, \eqref{percolbenad3_single}.  All simulation data are averages over 100 realisations of the growth of a network with $10^6$ nodes. The standard errors are smaller than the symbol sizes.\label{percol1}}
\end{figure}
In Fig.~\ref{percol1}, simulation data and various theoretical results for the time evolution of the size of the  giant component are plotted for the bachelor model. The dotted lines show the numerical solutions of the generating function theory without correlations, \eqref{percolbenad1_single}. Clearly, the results fail to provide a good approximation to the simulation data. The underlying reason is the assortative mixing present in the network that enlarges the number of links between hubs and thus enhances the formation of the largest cluster. Since \eqref{percolbenad1_single} neglects these correlations, it results in too small a giant cluster and too high a critical point. The approximation with nearest-neighbour correlations, \eqref{percolbenad2_single}, for which numerical results are shown with a dashed line, is better although it overcompensates the error of the approach without correlations. For $\alpha = 2$ and $\alpha = 8$, these theoretical results with  nearest-neighbour correlations suffice to give a reasonably accurate result, but for  $\alpha=3$ the results are not precise enough and the incorporation of next-nearest-neighbour correlations, as described by the system of Eqs.~(\ref{percolbenad3_single}), is necessary. Numerical solutions for this scheme are shown in Fig.~\ref{percol1} as solid lines. Clearly, these results match much better the simulations. The remaining deviations of this approach are related to the assortative next-next-nearest-neighbour correlations. The theory neglects them and predicts therefore a largest-cluster size which is slightly too small. Note moreover the alternation of the errors, caused by the alternation between assortative and disassortative correlations upon increase of the distance between the different nodes.

\begin{figure}[t]\centering
\includegraphics[angle=0,width = .5 \textwidth]{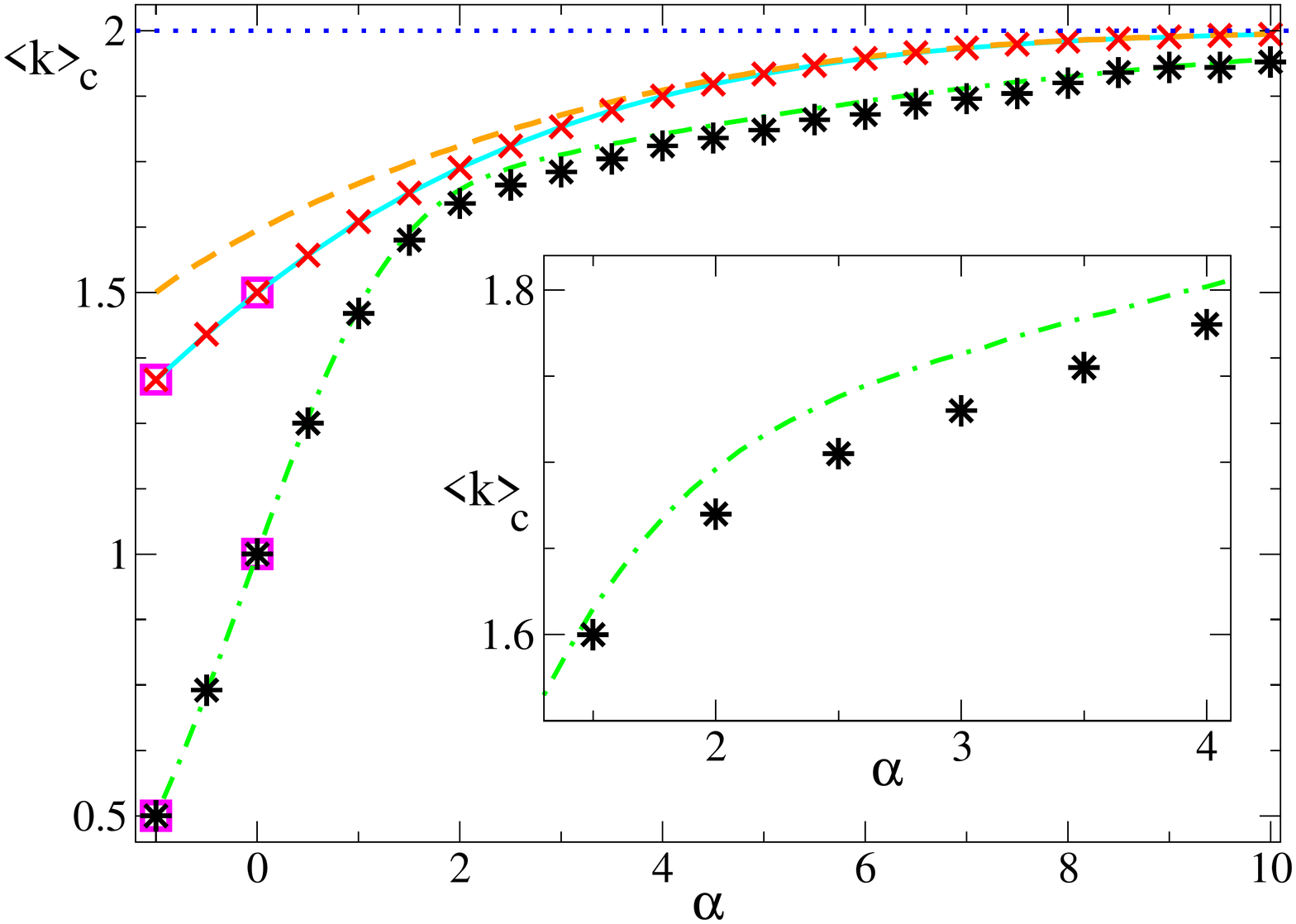}
\includegraphics[angle=0,width=.3\textwidth]{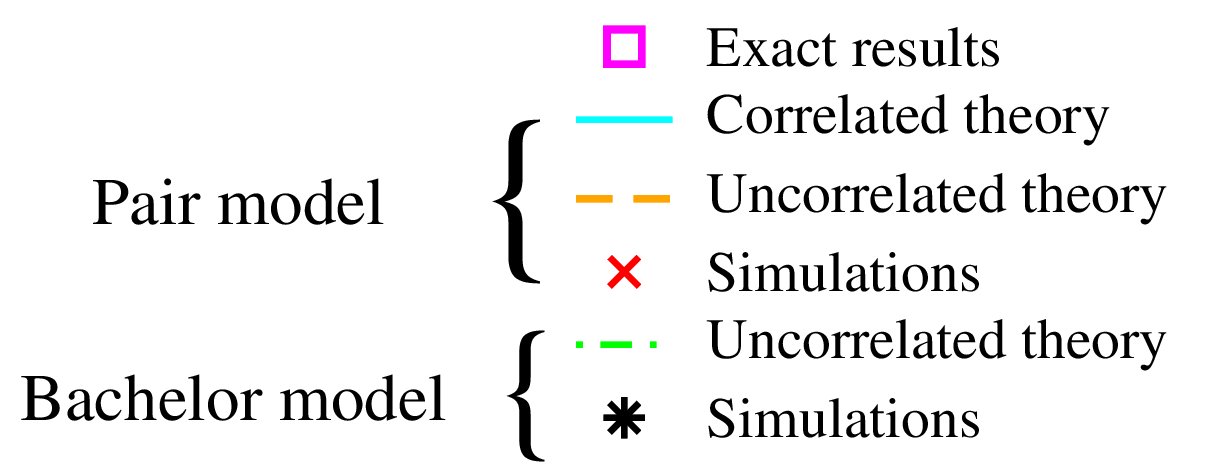}
\caption[Critical point of the percolation transition]{(Colour online) Critical point of the percolation transition as derived from simulations. The plot shows the critical mean degree $\avk_c$ in the network as a function of $\alpha$, both for the bachelor (black stars) and the pair (red crosses) model. The limiting value for $\alpha \rightarrow \infty$, $\avk_c =2$, is indicated with blue dots. For the \pairmodelpunt, numerical results to the relation \eqref{crittc} are shown with a cyan line. Furthermore, the plot visualises the effect of the correlations: interrupted lines indicate the critical point that can be found by solving numerically the uncorrelated generating functions theory, \eqref{percolbenad1_single}. These approximations are shown with an orange, dashed line (\pairmodelpunt) and a green, dot-dashed line (\bachelormodelpunt). Finally, some exact results are indicated with purple squares. The inset shows the data for the bachelor model in the regime $\alpha\in[1.5,4]$. All simulation data are averages over 100 realisations of the growth of a network with $10^6$ nodes. The standard errors are smaller than the symbol sizes.\label{tcbeide}
}
\end{figure}

The influence of the degree correlations on the percolation properties is also conspicuous in the results for the critical fraction, $\avk_c$. In Fig.~\ref{tcbeide}, the critical point observed in simulations is compared with a numerical solution to the uncorrelated generating functions scheme, \eqref{percolbenad1_single}. At first glance, the numerical results lie close to the simulation data for all values of $\alpha$. However, a detailed study in the regime $\alpha \in [2,4]$ shows that the approximation yields too high a critical fraction in this regime, as is expected from the reasoning above. These findings moreover illustrate the growing importance of the correlations as $\alpha$ increases from minus one to three and the decreasing importance as $\alpha$ further increases above four. 

We conclude that the percolation properties of the \bachelormodel can be found accurately within the generating functions framework. However, the inclusion of correlations is of vital importance. Depending on the value of $\alpha$ and the required accuracy, different types of correlations must be incorporated. Note, moreover, that for both model diversifications the clustering coefficient vanishes.

\subsection{Application to and modification for the \pairmodel}
In order to properly study the percolation properties of the \pairmodelpunt, we repeat that a distinction between two types of links has been made: the initial-time links, present at time zero, and the finite-time links, which are introduced later on.  Both types must be incorporated separately into a modified generating functions formalism. We therefore extend \eqref{percolbenad2_single} to include also the initial-link correlations:
\begin{subequations}\label{percolbenad2_pair}
\begin{align}
u^i_k &= \sum_q P^{i}_n(q|k) \left(u_q^{f}\right)^{q-1},\label{a}\\
u^f_k &=\sum_qP^{f}_n(q|k)u_q^{i}\left(u_q^{f}\right)^{q-2},\label{b}\\
\mathcal{S} &=1-\sum_q P^p(q) u_q^i\left(u_q^{f}\right)^{q-1},\label{c}
\end{align}
\end{subequations}
where $P_n^{i}(q|k)$ and $P_n^{f}(q|k)$ are the conditional \nearne probabilities, conditional on the fact that an initial-time link, respectively a finite-time link, has been followed. Both probabilities can be determined from \eqref{pair_corr} and \eqref{nfkq}:
\begin{subequations}
\begin{align}\label{condprob}
 P^{i}_n(q|k) &= P^p(q),\\
P_n^{f}(q|k)&=\frac{n_{k,q}^{f}}{(k-1)P^p(k)}.
\end{align}
\end{subequations}
 The proof of \eqref{percolbenad2_pair} can be found in Appendix \ref{appendixgenfuncinitialcorrel}.

Rather than solving these exact equations numerically, we again proceed analytically by neglecting the finite-time correlations. In that case, 
 approximation \eqref{nofinitetimecorrel} is valid and can be introduced into \eqref{percolbenad2_pair}. Some calculations (see Appendix \ref{appendixgenfuncinitialcorrel}) lead to
\begin{subequations}\label{percolbenad1_pair}
\begin{align}
u^f&=\frac{1}{\avk-1}\sum_{k,q}(q-1)P^p(k)P^p(q)\left(u^f\right)^{k+q-3},\\
\mathcal{S}&=1-\sum_{k,q} P^p(k)P^p(q) \left(u^f\right)^{k+q-2}.
\end{align}
\end{subequations}
Below the critical point, the sole solution is $u^f = 1$, which corresponds to a network in the absence of a giant cluster. An expansion around this critical value determines  the critical threshold $\avk_c$:
\begin{equation}
0 = \average{2}_c + \avk_c^2-6\avk_c+4,\label{crittc}
\end{equation}
where the subscript indicates that all quantities are evaluated at the critical point. Note that \eqref{crittc} results from the neglect of  correlations which are caused by the degree-dependent attachment. This assumption is nevertheless of reasonable validity since the percolation transition occurs when $\avk \leq 2$ and in Section \ref{sect_pair1} we argued that the finite-time correlations are negligible in this regime. 
\begin{figure}
\centering
\includegraphics[angle=0,width = .5 \textwidth]{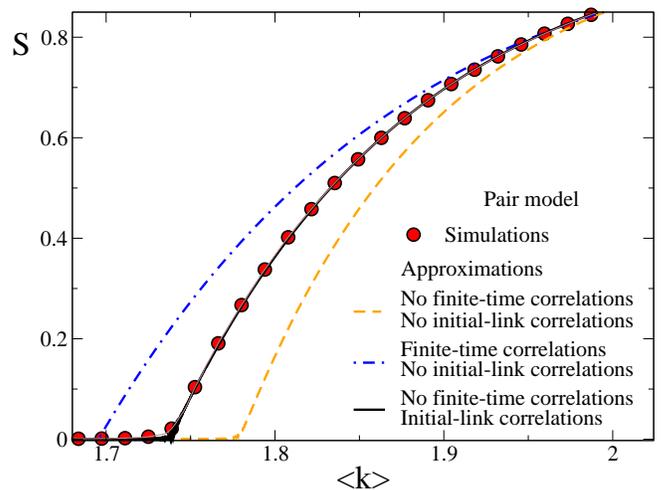}
\caption[The percolation transition for the \pairmodelpunt.]{(Colour online) The percolation transition for the \pairmodelpunt. The figure shows the fraction of the largest cluster as a function of the mean degree $\avk$ in the network. Various theoretical approximations (lines) are compared with simulation data (red circles) for the case $\alpha = 2$. The orange dashed line uses \eqref{percolbenad1_single} and ignores all correlations, while the blue dash-dotted line uses \eqref{percolbenad3_single} and therefore only incorporates the finite-time correlations. Lastly, the full line uses \eqref{percolbenad1_pair}, which does take the initial-link correlations into account but neglects the finite-time correlations. All simulation data are averages over 100 realisations of the growth of a network with $10^6$ nodes. The standard errors are smaller than the symbol sizes. \label{figpercol2}}
\end{figure}

In Fig.~\ref{tcbeide}, we compare numerical solutions to \eqref{crittc} with simulation data and indeed observe a good agreement between the approximate results and the simulations. Note, however, that the inclusion of initial-link correlations is very important. If, on top of the neglect of the finite-time correlations, also the initial-link correlations are discarded, the analytical results no longer match the simulation data. In Fig.~\ref{tcbeide}, we also show approximate results for the critical point in the \pairmodel in the absence of all types of correlations. These results are obtained by numerically solving the uncorrelated generating functions scheme, \eqref{percolbenad1_single}. Clearly, the uncorrelated critical points are inaccurate when $\alpha$ is small, which is exactly the regime in which the initial-link correlations are important. A similar conclusion can be drawn when the evolution of the size of the largest cluster is investigated. Fig.~\ref{figpercol2} compares simulation data for $\mathcal{S}$ with various approximations for the case $\alpha=2$. A first approximation (indicated with a dashed line) uses the results of the self-consistent system \eqref{percolbenad1_single} which ignores all kinds of correlations in the network. In the second approximation (shown with a dash-dotted curve), we use \eqref{percolbenad3_single}. In this approach, finite-time correlations are taken into account but the special initial condition associated with the initial-time links is neglected. Both approximations deviate significantly from the simulation results near the transition. 
In order to describe the initial-link correlations correctly, the modified generating functions scheme, \eqref{percolbenad1_pair}, must be used.   Although the finite-time correlations are neglected, numerical solutions to  \eqref{percolbenad1_pair} coincide with the simulation data.

We conclude that the percolation properties of the \pairmodel can be found with a modified generating functions theory which incorporates the initial-link correlations, but neglects the finite-time correlations.

\subsection{The bachelor and \pairmodel at criticality}
Simple heuristic arguments predict the dependence of the critical connectivity on $\alpha$. For very large and positive $\alpha$, nodes can only have one or two links, while, upon decrease of $\alpha$, non-zero densities appear for all degrees. Therefore  the percolation transition happens earlier in the linking procedure for smaller values of $\alpha$. Moreover, we expect to observe this behaviour qualitatively in both the bachelor and the pair model. Fig.~\ref{tcbeide} showed simulation results for the critical mean degree $\avk_c$ as a function of $\alpha$ for both models.
In the asymptotic limit $\alpha \rightarrow \infty$, the value $\avk_c=2$ is reached such that the percolation transition happens when all nodes have two links. For the \pairmodelpunt, $\avk_c$ increases gradually starting at  $\avk_c(\alpha = -1)=4/3$, while in the \bachelormodel the increase is much steeper, from $\avk_c(\alpha = -1) = 1/2$. The increase of $\avk_c$ for the \bachelormodel also follows a different trend: around $\alpha = 2$, a kink between two regimes is observed. 

Moreover, for all values of $\alpha$, the percolation transition occurs at a higher mean degree in the \pairmodel as compared to the \bachelormodelpunt. Due to its specific initial configuration, all clusters contain exactly two nodes when $\avk=1$ in the \pairmodelpunt, while in the \bachelormodel much larger structures already exist when $\avk=1$. The formation of a giant cluster is therefore delayed in the former, which explains the higher critical point. These results indicate that not only the linking probabilities, but also the initial condition of the growth process, greatly influence the percolation properties. 

In some special cases, the critical behaviour of the percolation transition can be studied analytically in more detail. Exact solutions for the generating functions formalism exist if $\alpha=0$ and $\alpha=-1$, both for the pair and the \bachelormodelpunt, while an approximation is constructed for large $\alpha$. Not only the critical point, but also the critical behaviour of the giant cluster is determined. The latter is studied using the critical exponent $\beta$ and critical constant $A$, defined using the relation $\mathcal{S}=A (t-t_c)^\beta$, which holds closely above the critical point.

\subsubsection{Special case: $\alpha=0$}
In the case of random link addition, there are no finite-time correlations in the network and \eqref{percolbenad1_single} therefore provides an exact expression for the size of the largest cluster for the bachelor model. Insertion of the exact degree distribution, \eqref{dda0}, into \eqref{percolbenad1_single} yields exact relations for the size of the giant cluster:
\begin{subequations}
\begin{align}
u^b(t)&=-\frac{1}{t}\textrm{LambertW}(-te^{-t}),\\
\mathcal{S}^b&= 1-e^{t(u^b-1)},\label{Salphanul}
\end{align}
\end{subequations}
where the LambertW-function is introduced \cite{olver2010}. Similar relations can be found for the pair model using the exact self-consistent set, \eqref{percolbenad2_pair}, and the exact degree distribution, \eqref{dda-1}. It is found that the percolation properties of both diversifications are trivially related by the relation $\mathcal{S}^b(2t)=\mathcal{S}^p(t)$, which also implies that the critical points are related by $\avk_c^b = 2\avk_c^p -2$.

The critical point is found by solving the requirement $u^b=1$, which yields
\begin{equation}
\avk_c^b = 2\avk_c^p -2 = 1,
\end{equation}
which matches for the bachelor model the well-known results for the \erdosrenyi graph \cite{erdos1960,bollobas2001}.
Note that both critical points are confirmed by simulations, as can be seen in Fig.~\ref{tcbeide}.  For both diversifications, the critical behaviour can be found by expanding the self-consistent equations about the critical point. One obtains
\begin{subequations}
\begin{align}
\mathcal{S}^b &\sim 2(\avk-1),\\
\mathcal{S}^p&\sim 4\left(\avk-\frac{3}{2}\right).\label{analytischeresultaten}
\end{align}
\end{subequations}
In both cases, the critical exponent $\beta$ thus equals one, while the critical constant $A$ depends on the model. The results for the \bachelormodel match once again the well-known expressions for the Erd\H{o}s-R\'{e}nyi graph \cite{bollobas2001}.
The validity of these analytical findings for the critical behaviour is substantiated by a numerical evaluation of the generating functions theory. Close to the critical point, the results for the size of the largest cluster can be fitted to $\mathcal{S}=A(t-t_c)^\beta$.  For instance, performing the numerical scheme for the \pairmodel yields
\begin{subequations}
\begin{align}
\avk_c^p \isarray 1.498 \pm 0.002,\\
\beta \isarray 1.01 \pm 0.02,\\
A \isarray 3.92 \pm 0.10,
\end{align}
\end{subequations}
which agrees, within the error margins, with the analytical results in \eqref{analytischeresultaten}.

\subsubsection{Special case: $\alpha=-1$}
Finite-time correlations are also absent in the preferential-attachment limit, i.e., when $\alpha = -1$. \eqref{percolbenad1_single} and \eqref{percolbenad1_pair} again provide exact expressions for the size of the largest cluster in the bachelor, respectively~the pair model. Using the exact degree distributions, \eqref{dda-1}, the self-consistent equations for $\mathcal{S}$ can be solved analytically for both diversifications. For the bachelor model, one for instance obtains
\begin{equation}\label{Salpha-1bach}
\mathcal{S}^b = \left\{
\begin{array}{ccc}
0&\textrm{if}&t<\dfrac{1}{2}\\
\dfrac{t-2+\sqrt{t(4+t)}}{t+\sqrt{t(4+t)}}&\textrm{if}&t>\dfrac{1}{2}
\end{array} \right..
\end{equation}
The critical point $\avk_c^b=1/2$ agrees with simulation data, as can be seen in Fig.~\ref{tcbeide}. An expansion of \eqref{Salpha-1bach} about the critical point yields
\begin{equation}
\mathcal{S}^b \sim \frac{8}{3}\left(\avk-\frac{1}{2}\right).
\end{equation}
A similar calculation is performed for the \pairmodelpunt. The final results are
\begin{subequations}
\begin{align}
\avk_c^p\isarray \frac{4}{3},\\
\mathcal{S}^p&\sim 3\left(\avk-\frac{4}{3}\right).
\end{align}
\end{subequations}
The critical point is again supported by simulation data, as was once more shown in Fig.~\ref{tcbeide}. Solutions to the rate equation for the degree distribution and the generating functions theory also confirm the critical behaviour. Numerically solving the generating functions theory leads to the results
\begin{subequations}
\begin{align}
\avk^p_c \isarray 1.33 \pm 0.01,\\
\beta \isarray 0.99 \pm 0.03,\\
A \isarray 2.93 \pm 0.15,
\end{align}
\end{subequations}
which match, within the error margins, the analytical results.

\subsubsection{Special case: large $\alpha$}
In the limit $\alpha = +\infty$, new links are always laid between (two of) the nodes that have the lowest connectivities. If $1\leq\avk \leq 2$, the evolution of the cluster-size distribution is then equivalent to that in a process in which new links are always laid between two randomly chosen clusters. Consequently, also the percolation properties match with those in such a process. In Refs.~\cite{cho2010,manna2011} it was shown analytically and using simulations that this specific random cluster-linking process gives rise to an explosive, first-order percolation transition \footnote{Note, however, that the initial conditions in Ref.~\cite{cho2010} do not match those in our pair model. It can, however, be shown analytically that this does not affect the occurrence of a first-order transition.}. In this section, we study the transition to the explosive limiting situation using an expansion for large $\alpha$.
Recall moreover that in the limit $\alpha \rightarrow +\infty$, the bachelor and the pair models are equal if $\avk \geq 1$. We therefore limit our studies for large $\alpha$ to the percolation properties of the pair model and anticipate the differences with the bachelor model to be small.

\begin{figure}
\centering
\includegraphics[angle=0,width = .45 \textwidth]{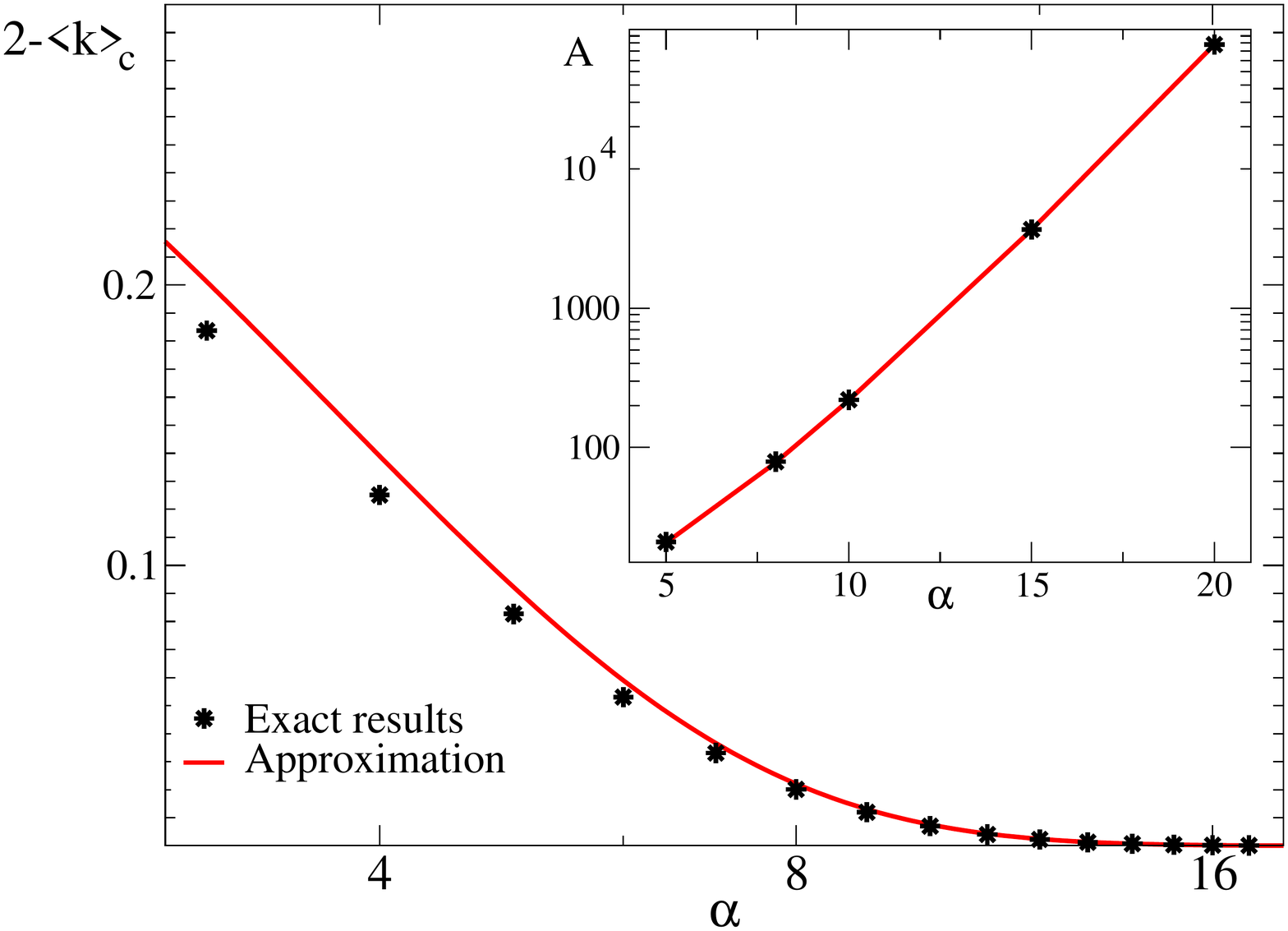}
\caption[Approximation for the critical  connectivities  for large values of $\alpha$.]{(Colour online) Comparison of the approximation for large $\alpha$ for the critical behaviour of the percolation transition with numerical results from the generating functions theory. The main figure shows results for the critical point $2-\avk_c$  of the pair model in a semi-logarithmic scale. The full red line is the solution to the approximation without correlations, \eqref{approx1}, while the black stars indicate the numerical solutions of \eqref{percolbenad1_pair}. In the inset, the amplitude $A$ of the leading singularity $\mathcal{S}=A(t-t_c)^\beta$ about the critical point is shown as a function of $\alpha$. The full red line represents the approximation, \eqref{benadA1}, while the black stars indicate the results of a fit $\mathcal{S}=A(t-t_c)$ to the numerical solution of the generating functions theory.
\label{tcbenad}}
\end{figure}

The starting point of the study of the percolation transition is the expanded degree distribution for large $\alpha$, Eq.~(\ref{degree_alpha_infin}), which is valid as long as $(3/2)^{-\alpha}\ll 1$ and $|\twee \ln(1-t)| \ll 1$, where, as before, $\epsilon = 2^{-\alpha}$. The finite-time correlations are once more neglected, since they vanish for $\alpha \rightarrow\infty$. We also neglect the initial-link correlations, since simulations indicate that their influence on the critical point also vanishes for large $\alpha$ \footnote{We could as well include the initial-link correlations, thus expanding the self-consistent set (\ref{percolbenad1_single}) for large $\alpha$. However, due to a non-trivial cancellation of errors, the uncorrelated calculation outperforms this scheme with initial-link correlations, as is shown in Ref.~\cite{phdthesis}. Therefore, we will henceforth neglect the initial-link correlations.}.  Inserting the approximate degree distribution in the uncorrelated generating functions scheme, \eqref{percolbenad1_single}, yields an approximate criterion for the critical time $t_c$:
\begin{equation}\label{approx1}
1-t_c=-2\twee [t_c + \ln(1-t_c)].
\end{equation}
Using the LambertW-function \cite{olver2010}, this implicit equation can be solved for the critical point:
\begin{equation}\label{hierboven}
t_c = 1 - \frac{2\twee}{1-2\twee}\textrm{LambertW}\left(\frac{1-2\twee}{2e\twee}\right).
\end{equation} 
Note that this expression is only meaningful in case the large $\alpha$-expansion is still valid at the critical point, i.e., if the obtained critical time satisfies $|\twee \ln(1-t_c)| \ll 1$ for large $\alpha$. A numerical evaluation indicates that the factor $|\twee\ln(1-t_c)|$ decreases with increasing $\alpha$. We obtain $0.0745$ for $\alpha =5$, while for $\alpha = 10$, we obtain $0.0048$. The expanded degree distributions can thus be used at the critical point for large $\alpha$.
In Fig.~\ref{tcbenad}, the approximate critical points $\avk_{c,approx}$, obtained using \eqref{hierboven}, are compared with numerical solutions to the exact self-consistent equations $\avk_{c,\textrm{exact}}$. Clearly, the approximation improves significantly upon increase of $\alpha$. Its accuracy  can be quantified by means of the parameter $\gamma$ which is defined as follows:
\begin{equation}
\gamma = 1 - \frac{2-\avk_{c,\textrm{appr}}}{2-\avk_{c,\textrm{exact}}}.
\end{equation} 
Simulations indicate that $\gamma = 4.9\cdot 10^{-3}$ for $\alpha = 5$, while $\gamma$ further decreases to $\gamma = 2.9\cdot 10^{-4}$ for $\alpha = 10$. We conclude that the expansion for large $\alpha$ provides an accurate approximation for the critical point if $\alpha > 5$.

Within the large-$\alpha$ approximation, the critical exponent $\beta$ and the critical constant $A$ can be determined. We define $\xi\equiv t-t_c$ and work in the regime $\xi \ll 1$. Expanding the generating functions theory for large $\alpha$ in $\xi$ and retaining only the lowest order yields  $\mathcal{S} \sim A\xi^\beta$, with $\beta =1$ and
\begin{align}
A&=\frac{2}{3}\frac{1+t_c}{1-t_c}\left(1+\frac{2\twee t_c}{1-t_c}\right).\label{A1}
\end{align}
Insertion of the critical point, \eqref{hierboven}, provides a numerical result for $A$, while an analytical approximation is obtained by an expansion in a Laurent series. Retaining only the leading two terms of the series,
\begin{widetext}
\begin{equation}\label{benadA1}
A \approx \frac{2}{3\twee}\left(1+\frac{1}{\textrm{LambertW}\left(\frac{1}{2e\twee}\right)}\right)-\frac{2}{3}\left(3+\frac{1}{\textrm{LambertW}\left(\frac{1}{2e\twee}\right)}\right)
\end{equation}
\end{widetext}
is obtained. Since $\textrm{LambertW}(x)$ diverges for $x\rightarrow +\infty$ \cite{olver2010}, \eqref{benadA1} straightforwardly implies that  $A$ diverges for $\epsilon\rightarrow 0$ and thus for $\alpha \rightarrow +\infty$.
The validity of this expansion for large $\alpha$ is again supported by a numerical evaluation of the exact generating functions theory. Close to the critical point, the numerical results for the size of the giant cluster can be fitted to $\mathcal{S} = A(t-t_c)^\beta$. Within the error margins, the critical exponent $\beta$ is equal to one for all values of $\alpha$, while the inset of Fig.~\ref{tcbenad} compares the estimate for $A$ with the value  obtained in the fit.  The approximate result is found to agree with the theoretical results for $\alpha > 5$.

The results indicate how the transition from the normal to the explosive percolation regime occurs in our model. The critical exponent $\beta$ equals one for all finite values of $\alpha$. Surprisingly, the critical amplitude $A$ increases rapidly as the explosive regime is approached. The transition to the explosive percolation for $\alpha \rightarrow+\infty$ is thus not caused by a vanishing critical exponent $\beta$, but by a diverging critical amplitude $A$.

\section{Conclusions}
We have introduced a network growth model for studying the growth of a network with a fixed number of nodes using general degree-dependent linking probabilities. We have performed a detailed study of the construction process and the constructed networks. Rate equations determine the time evolution of the degree distribution and the degree correlations, and a generating functions formalism provides accurate results for the percolation properties, although the incorporation of degree correlations in the theory is of vital importance. 

To illustrate our findings using simulations, we have introduced two diversifications which provide a cross-over from a preferential attachment rule to a process featuring explosive percolation: the bachelor model, with initially a set of unconnected nodes, and the pair model, in which nodes are initially coupled into pairs. In both variants, a single parameter $\alpha$ determines the dynamics. If $\alpha$ is negative, links are preferably laid between nodes with a larger degree, while the reverse statement holds for positive $\alpha$.  If $\alpha$ vanishes, the model mimics the network growth of the Erd\H{o}s-R\'{e}nyi random network, while $\alpha =-1$ corresponds to the preferential attachment of the Barab\'{a}si-Albert model. In the limit $\alpha \rightarrow \infty$, an explosive percolation transition is retrieved.

For our two diversifications, numerical solutions to the rate equations are in accord with the outcome of  Monte-Carlo simulations. The results furthermore show that power-law degree distributions are never obtained in our model, which extends the finding of \barabasi \textit{et al.}~that only processes with an increasing number of nodes facilitate the scale-free structure~\cite{barabasi1999}.  Although the degree distributions of both diversifications are trivially related to one another, the different initial conditions cause major distinctions for the degree correlations in the constructed networks. While in the bachelor model correlations only appear as a consequence of the linking process, the network is already highly correlated in the initial stage if the pair model is used. The two diversifications also illustrate the influence of degree correlations on the percolation properties. In the pair model, it suffices to include the initial-link correlations, since they are dominant at criticality, while,  depending on the value of $\alpha$ and the required accuracy, different types of finite-time correlations must be incorporated for the bachelor model.

We have paid special attention to three qualitatively different special cases: $\alpha = 0$ (random network growth), $\alpha =-1$ (preferential attachment) and $\alpha \rightarrow +\infty$ (explosive percolation). For large $\alpha$, the construction process is studied using approximate degree distributions and expanded generating functions, which indicate that the transition to the explosive regime is marked by a rapid increase of the critical amplitude, while the critical exponent $\beta$ remains unchanged and equal to one for all finite values of $\alpha$.

Although the rate equation and generating functions approach have here only been applied to the specific linking probabilities $p_k \propto k^{-\alpha}$ and $p_k\propto (k+1)^{-\alpha}$, the techniques presented in the paper are of general validity for all degree-dependent linking probabilities. We leave the application of the methods to other network growth models as a challenge for future research. We refer to Refs.~\cite{havlin2012} and~\cite{sahini1994} for applications of percolation to different fields of science.

\appendix

\section*{Acknowledgement}
H.H. acknowledges support of the FWO (Fonds voor Wetenschappelijk Onderzoek - Vlaanderen). J.O.I. is supported by KU Leuven Research Grant OT/11/063.

\section{Derivation of the rate equation for the degree distribution\label{appendixrateeqdd}}
In this appendix, the mean-field rate equation for the degree distribution,
\begin{equation}\label{samappendix}
\frac{d}{dt}P(k) = p_{k-1}P(k-1) -  p_kP(k),
\end{equation}
will be proven to hold in the thermodynamic limit $N\rightarrow\infty$. The derivation describes the evolution of the number of nodes with $k$ links, $N_k = NP(k)$. The increment of $N_k$ in a single time-step is denoted as
\begin{equation}
\Delta N_k(t) \equiv N_k(t+\Delta t) - N_k(t),
\end{equation}
where $\Delta t$ is the duration of a time-step.
If we take into account the fact that two nodes are chosen at the same time, different cases will contribute to  $\Delta N_k$. In the remainder of the section, we will denote the case in which nodes with degrees $k_1$ and $k_2$ are selected as $[k_1,k_2]$. Note that the behaviour of $[k_1,k_2]$ and $[k_2,k_1]$ is similar, therefore both will be considered simultaneously, which will induce the introduction of factors of two. The five possible cases contributing to $\Delta N_k$ are 
\begin{enumerate}
\item $[k,x]$\qquad \quad with $x\neq k-1,k$
\item $[k-1,x]$ \quad with $x\neq k-1,k$
\item $[k-1,k]$
\item $[k,k]$
\item $[k-1,k-1]$
\end{enumerate}
The selection probability and the change in $N_k$ are now determined for the different possibilities. Recall that the probability to select a single node with degree $k$ is $p_k/N$. The probability to select \textit{a} node with degree $k$ is thus $p_kP(k)$.

The selection probability of case 1 is
\begin{equation}
p^{(1)} = 2p_kP(k)(1-p_kP(k)-p_{k-1}P(k-1)),
\end{equation}
since the probability to select the node with degree $k$ is $p_kP(k)$ and the probability to select the node with a degree different from $k$ and $k-1$ is $(1-p_kP(k)-p_{k-1}P(k-1))$.  The factor 2 appears since $[k,x]$ and $[x,k]$ are distinguishable. If this case occurs, the number of nodes with degree $k$ diminishes by one. The change in $N_k$ in the first case is thus
\begin{equation}
\Delta N_k^{(1)} = -2p_kP(k)(1-p_kP(k)-p_{k-1}P(k-1)).
\end{equation}
Similarly, the second case contributes
\begin{equation}
\Delta N_k^{(2)} = +2p_{k-1}P(k-1)(1-p_kP(k)-p_{k-1}P(k-1))
\end{equation}
to $\Delta N_k$. The third case actually does not contribute to $\Delta N_k$, since there is no net change in the number of nodes with degree $k$. In the fourth case, the selection probability amounts to
\begin{equation}
p^{(4)} = p_k^2P(k)^2,
\end{equation}
without a factor of two since the two possibilities are indistinguishable in this case. The number of nodes with degree $k$ diminishes by two if two nodes with degree $k$ are selected. Therefore,
\begin{equation}
\Delta N_k^{(4)} = -2(p_kP(k))^2.
\end{equation}
By the same token, for case 5,
\begin{equation}
\Delta N_k^{(5)} = +2(p_{k-1}P(k-1))^2.
\end{equation}
The total change in $N_k$ in a single time-step is now
\begin{subequations}
\begin{align}
\Delta N_k &= \Delta N_k^{(1)}+\Delta N_k^{(2)}+\Delta N_k^{(4)}+\Delta N_k^{(5)}\\
&=2(p_{k-1}P(k-1)-p_kP(k)),
\end{align}
\end{subequations}
and, consequently,
\begin{subequations}
\begin{align}
\Delta P(k) &= P(k,t+\Delta t) - P(k,t)\\
&=\frac{2}{N}(p_{k-1}P(k-1)-p_kP(k)).
\end{align}
\end{subequations}
Since the time increases by $2/N$ every time-step, i.e., $\Delta t = 2/N$, taking the limit $N\rightarrow \infty$ implies
\begin{equation}
\frac{\Delta P(k)}{2/N}\rightarrow \frac{\partial}{\partial t}P(k),
\end{equation}
which leads straightforwardly to \eqref{samappendix}.

\section{Derivation of the rate equation for the degree correlations\label{appendixrateeqnn}}
In this Appendix, we show how simple arguments lead to the mean-field rate equation for \nearne degree correlations,
\begin{align}\label{badalona}
\frac{dn_{k,q}}{dt} &= p_{k-1}n_{k-1,q}+p_{q-1}n_{k,q-1}-n_{k,q}\left(p_k + p_q\right) \nonumber\\&\quad\quad +p_{k-1}p_{q-1}P(k-1)P(q-1),
\end{align}
where all quantities are defined as in Section \ref{sectiondegreedegreecorrelations}. The key quantity in the derivation is $N_{k,q}=Nn_{k,q}$, which is the number of links between nodes with degrees $k$ and $q$ if $k\neq q$, while $N_{k,k}=Nn_{k,k}$ represents  twice the number of links between two nodes with degree $k$. Due to this deviating definition in case $k=q$, this special case must be treated separately. In the derivation, we describe the evolution of $N_{k,q}$ in a single time-step and therefore introduce its variation
\begin{equation}
\Delta N_{k,q}(t) \equiv N_{k,q}(t+\Delta t) - N_{k,q}(t),
\end{equation}
where $\Delta t$ is the duration of a time-step.

\subsection{Derivation of the rate equation for $N_{k,q}$ with $k\neq q$}
If we take into account the fact that two nodes are chosen simultaneously, different cases will contribute to  $\Delta N_{k,q}$ if $k\neq q$. We denote the case in which nodes with degrees $k_1$ and $k_2$ are selected as $[k_1,k_2]$ and  consider the behaviour of the case in which $[k_2,k_1]$ is chosen at the same time, which will induce the introduction of factors of two. The eight possible cases in which  $\Delta N_{k,q}$ changes, are then
\begin{enumerate}
\item $[k,x]$\qquad\quad with $x\neq q-1,q$
\item $[x,q]$\qquad\quad with $x\neq k-1,k$
\item $[k-1,x]$\quad with $x\neq q-1,q$
\item $[x,q-1]$\quad with $x\neq k-1,k$
\item $[k,q]$
\item $[k-1,q]$
\item $[k,q-1]$
\item $[k-1,q-1]$
\end{enumerate}
The selection probability and the change in $N_{k,q}$ are  now determined for the different possibilities. The selection probability of case 1 is
\begin{equation}
p^{(1)} = 2p_kP(k)(1-p_qP(q)-p_{q-1}P(q-1)),
\end{equation}
where the factor two indicates that $[k,q]$ and $[q,k]$ are indistinguishable. In a mean-field approximation, $N_{k,q}$ decreases with the average number of links attached to a node with degree $k$ which leads to a node with degree $q$. Since each of the $k$ links attached to the $k$-node end at a $q$-node with probability $P_n(q|k)$, the total change in $N_{k,q}$ is 
\begin{equation}
\Delta N_{k,q}^{(1)} = -2p_kn_{k,q}(1-p_qP(q)-p_{q-1}P(q-1)),
\end{equation}
where we also used that $n_{k,q}=kP(k)P_n(q|k)$, see \eqref{Nkq}. Similar calculations yield the change in the cases 2 to 7:
\begin{subequations}
\begin{align}
\Delta N_{k,q}^{(2)} &= -2p_qn_{k,q}(1-p_kP(k)-p_{k-1}P(k-1)),\\
\Delta N_{k,q}^{(3)} &= +2p_{k-1}n_{k-1,q}(1-p_qP(q)-p_{q-1}P(q-1)),\\
\Delta N_{k,q}^{(4)}&=+2p_{q-1}n_{k,q-1}(1-p_kP(k)-p_{k-1}P(k-1)),\\
\Delta N_{k,q}^{(5)}&=-2p_{k}p_{q}n_{k,q}(P(q)+P(k)),\displaybreak[0]\\
\Delta N_{k,q}^{(6)}&=2p_{k-1}p_{q}(P(q)n_{k-1,q}-P(k-1)n_{k,q}),\\
\Delta N_{k,q}^{(7)}&=2p_{k}p_{q-1}(P(k)n_{k,q-1}-P(q-1)n_{k,q}).
\end{align}
\end{subequations}
In case 8, the derivation changes slightly because the insertion of a link between a node with degree $k-1$ and one with degree $q-1$ causes the introduction of an extra link in the set of links between nodes with degrees $k$ and $q$. Therefore,
\begin{align}
\Delta N_{k,q}^{(8)}&=2p_{k-1}p_{q-1}\Big(P(k-1)n_{k,q-1}\\&+P(q-1)n_{k-1,q}+P(k-1)P(q-1)\Big).\nonumber
\end{align}
Adding all eight terms yields the total change  in $N_{k,q}$: 
\begin{subequations}
\begin{align}
\Delta N_{k,q} &= 2\Big(p_{k-1}n_{k-1,q}+p_{q-1}n_{k,q-1}-n_{k,q}\left(p_k + p_q\right) \nonumber\\&\quad\quad +p_{k-1}p_{q-1}P(k-1)P(q-1)\Big).
\end{align}
\end{subequations}
Since the time increases by $2/N$ every time-step, taking the limit $N\rightarrow \infty$ implies
\begin{equation}
\frac{\Delta N_{k,q}}{2}\rightarrow \frac{\partial}{\partial t}n_{k,q},
\end{equation}
which proves the correctness of \eqref{badalona} if $k\neq q$.

\subsection{Derivation of the rate equation for $N_{k,k}$}
The derivation of a rate equation for $n_{k,k}$ proceeds analogously, but we have to keep in mind that the definition of $n_{k,k}$ and $N_{k,k}$ differs since the latter quantity represents \textit{twice} the number of links between two nodes with degrees $k$. To facilitate the notation, we introduce $\tilde{N}_{k,q}\equiv N_{k,q}/2$. A change in $\tilde{N}_{k,k}$ is only observed in one of the following cases:
\begin{enumerate}
\item $[k,x]$\qquad\quad with $x\neq k-1,k$
\item $[k-1,x]$\quad with  $x\neq k-1,k$
\item $[k-1,k]$
\item $[k-1,k-1]$
\item $[k,k]$
\end{enumerate}
Since $\tilde{N}_{k,k}$ represents the number of links, its alteration  for the first three cases is exactly the same as for the corresponding cases in the previous section, i.e.,
\begin{subequations}
\begin{align}
\Delta \tilde{N}_{k,k}^{(1)} &= -2p_kn_{k,k}(1-p_kP(k)-p_{k-1}P(k-1)),\\
\Delta \tilde{N}_{k,k}^{(2)} &= +2p_{k-1}n_{k,k-1}(1-p_kP(k)-p_{k-1}P(k-1)),\\
\Delta \tilde{N}_{k,k}^{(3)} &= 2p_{k-1}p_k(n_{k,k-1}P(k)-P(k-1)n_{k,k}).
\end{align}
\end{subequations}
In the last two cases, caution is in order, since the two possibilities are now indistinguishable. Of course, the links attached to  \textit{both} endpoints of the new link must be counted in the change of $\tilde{N}_{k,k}$, which yields
\begin{subequations}
\begin{align}
\Delta \tilde{N}_{k,k}^{(4)} &= -2(p_k)^2P(k)n_{k,k},\\
\Delta \tilde{N}_{k,k}^{(5)} &= 2(p_{k-1})^2P(k-1)n_{k,k-1}+(p_{k-1}P(k-1))^2,
\end{align}
\end{subequations}
where the last term again stems from the addition of the extra link. A back-of-the-envelope calculation yields
\begin{equation}
\Delta \tilde{N}_{k,k} = 2(-p_kn_{k,k}+p_{k-1}n_{k,k-1})+(p_{k-1}P(k-1))^2.
\end{equation}
Since the time increases by $2/N$ every time-step, taking the limit $N\rightarrow \infty$ implies
\begin{equation}
\Delta \tilde{N}_{k,q}\rightarrow \frac{\partial}{\partial t}n_{k,q}.
\end{equation}
Thus
\begin{equation}
\frac{\partial}{\partial t}n_{k,q}=2(-p_kn_{k,k}+p_{k-1}n_{k,k-1})+(p_{k-1}P(k-1))^2,
\end{equation}
which is exactly the $k=q$ limit of \eqref{badalona}. We conclude that we have proven the validity of this rate equation for all possible cases.

\section{Derivation of the generating functions theory in the presence of initial-link correlations \label{appendixgenfuncinitialcorrel}}
In our pair model, the network starts from an initial state in which all nodes are combined into pairs.  Here, we derive the self-consistent equations for the size of the largest cluster in the presence of initial-time links~(see \eqref{percolbenad2_pair}), 
\begin{subequations}\label{percolbenad2_pairapp}
\begin{align}
u^i_k &= \sum_q P^{i}_n(q|k) \left(u_q^{f}\right)^{q-1},\label{aapp}\\
u^f_k &=\sum_qP^{f}_n(q|k)u_q^{i}\left(u_q^{f}\right)^{q-2},\label{bapp}\\
\mathcal{S} &=1-\sum_q P^p(q) u_q^i\left(u_q^{f}\right)^{q-1},\label{capp}
\end{align}
\end{subequations}
and its modified form (see \eqref{percolbenad1_pair})
\begin{subequations}
\begin{align}\label{percolbenad1_pairapp}
u^f&=\frac{1}{\avk-1}\sum_{k,q}(q-1)P^p(k)P^p(q)\left(u^f\right)^{k+q-3},\\
\mathcal{S}&=1-\sum_{k,q} P^p(k)P^p(q) \left(u^f\right)^{k+q-2},
\end{align}
\end{subequations}
which is valid if finite-time correlations are neglected. 
The derivation is based on the generating functions theory for the percolation problem, which is discussed in detail in Refs.~\cite{dorogovtsev2007,newman2001}. In order not to complicate the notation unnecessarily, the derivation will, however, be presented using probabilities rather than generating functions. The full derivation using generating functions can be found in Ref.~\cite{phdthesis}.

The correlations are described using the conditional nearest-neighbour degree probabilities $P_n^{i}(k|q)$ and $P_n^{f}(k|q)$, conditional on the fact that an initial-time link, respectively a finite-time link, has been followed. We furthermore introduce the probability  $\tilde{u}_k$ that a random node with degree $k$ belongs to a finite cluster of any size and the probabilities  $u^{i}_q$ and $u^{f}_q$ that an initial-time, respectively a finite-time link, originating at a node of degree $q$ leads to a finite cluster of any size at the other end. Relations between those quantities can be found easily, if we assume that the finite clusters in the network are tree-like. For instance, a node can only belong to a finite cluster, if all its links lead to a finite cluster. If we keep in mind that a node with degree $k$ has a single initial-time link, while all other $k-1$ links are laid at finite time, the probability that a random node with degree $k$ belongs to a finite cluster is then given by 
\begin{equation}
\tilde{u}_k =  u_k^i\left(u_k^{f}\right)^{k-1}.
\end{equation}
The probability that a randomly chosen node (of any degree) belongs to a finite cluster, henceforth denoted with $\tilde{u}$, is then
\begin{align}
\tilde{u} = \sum_kP(k)u_k^i\left(u_k^{f}\right)^{k-1}.
\end{align}
The quantity $\tilde{u}$ is simply related to the size of the giant cluster, since $\tilde{u} +\mathcal{S} =1$, because nodes either belong to a finite cluster or to the giant cluster. Therefore, 
\begin{align}
\mathcal{S} =1- \sum_kP(k)u_k^i\left(u_k^{f}\right)^{k-1},
\end{align}
which is the third line of \eqref{percolbenad2_pairapp}. With a similar reasoning, the first two lines of  \eqref{percolbenad2_pairapp} can be deduced. 

Note that the self-consistent system of \eqref{percolbenad2_pairapp} can be reduced to a system with only two equations by inserting \eqref{aapp} in \eqref{bapp} and \eqref{capp}:
\begin{subequations}
\begin{align}
u^f_k &=\sum_{q,s}  P^{i}_n(s|q) P^{f}_n(q|k)   \left(u_s^{f}\right)^{s-1}\left(u_q^{f}\right)^{q-2},\\
\mathcal{S} &=1-\sum_{q,k} P(q)   P^{i}_n(k|q) \left(u_k^{f}\right)^{k-1}\left(u_q^{f}\right)^{q-1}.
\end{align}
\end{subequations}
Insertion of the expressions for the conditional probabilities, \eqref{condprob}, yields
\begin{subequations}\label{percolbenad2_pairbis}
\begin{align}
u^f_k&=\sum_{q,s}\frac{n_{k,q}^{f}}{(k-1)P^p(k)} P^p(s)\left(u^f_s\right)^{s-1}\left(u^f_q\right)^{q-2},\\
\mathcal{S}&=1-\sum_{q,k} P^p(k)P^p(q)\left(u^f_k\right)^{k-1}\left(u^f_q\right)^{q-1}.
\end{align}
\end{subequations}
In  case the finite-time correlations can be neglected,  approximation \eqref{nofinitetimecorrel} is valid and can furthermore be introduced into \eqref{percolbenad2_pairbis}, which yields \eqref{percolbenad1_pairapp}.

\bibliography{bib}
\end{document}